\documentclass[12pt, draftclsnofoot, onecolumn]{IEEEtran}
\ifCLASSINFOpdf
\usepackage[pdftex]{graphicx}
\else
\fi
\usepackage[cmex10]{amsmath}
\usepackage{url}
\usepackage{cite}
\usepackage{subcaption}

\usepackage{algorithm}
\usepackage{algpseudocode}
\usepackage{tikz}
\usetikzlibrary{shapes,arrows,chains}

\usepackage{enumitem}
\setlist[enumerate,1]{leftmargin=15pt}

\begin{document}

\tikzset{
	decision/.style={draw,diamond,fill=blue!20, inner sep=1pt, align=center,aspect=2, minimum
	height=2em},
	block/.style={draw, rectangle, fill=blue!20, rounded corners, inner sep=2pt, align=center, minimum
	height=2em},
	line/.style={draw, -latex},
	cloud/.style={draw, ellipse, fill=red!20, node distance=3cm, minimum height=2em}
}%

%
\title{An Energy-Aware WLAN Discovery Scheme for LTE HetNet\IEEEauthorrefmark{2}\thanks{\IEEEauthorrefmark{2}This paper is a substantially expanded and modified version of the work in \cite{Wcnc}.}}
%
%
%

\author{Pravjyot~Singh~Deogun\IEEEauthorrefmark{1}, 
        Arghyadip~Roy\IEEEauthorrefmark{7},~\IEEEmembership{Student Member,~IEEE,}
        Abhay~Karandikar\IEEEauthorrefmark{7},~\IEEEmembership{Member,~IEEE}
\thanks{\IEEEauthorrefmark{1}P. S. Deogun is with Samsung R\&D Institute, Bangalore-560037, India, e-mail: ps.deogun@samsung.com. This work was performed when the author was working at IIT Bombay.}%
\thanks{\IEEEauthorrefmark{7}A. Roy and A. Karandikar are with the Department
of Electrical Engineering, Indian Institute of Technology Bombay,
Mumbai-400076, India, e-mail: $\lbrace$arghyadip,karandi$\rbrace$@ee.iitb.ac.in.}}
\maketitle

\begin{abstract}
	Recently, there has been significant interest in the integration and co-existence of Third Generation Partnership Project (3GPP) 
	Long Term Evolution (LTE) with other Radio Access Technologies,
	like IEEE 802.11 Wireless Local Area Networks (WLANs). Although, the inter-working of IEEE 802.11 WLANs with 3GPP
	LTE has indicated enhanced network performance in the context of capacity and load balancing, the WLAN discovery scheme
	implemented in most of the commercially available smartphones is very inefficient and results in high battery drainage.
	In this paper, we have proposed an energy efficient WLAN discovery scheme for 3GPP LTE and IEEE 802.11 WLAN inter-working scenario.
	User Equipment (UE), in the proposed scheme, uses 3GPP network assistance along with
	the results of past channel scans, to optimally select the next channels to scan. Further, we have also
	developed an algorithm to accurately estimate the UE's mobility state, using 3GPP network signal strength
	patterns. We have implemented various discovery schemes in Android framework, to evaluate the performance of our proposed scheme 
	against other solutions in the literature. Since, Android does not support selective scanning
	mode, 
	we have implemented
	modules in Android to enable selective scanning.
	Further, we have also used simulation studies and justified the results using power consumption modeling.
	The results from the field experiments and simulations have shown high power savings using the proposed scanning scheme
	without any discovery performance deterioration.
\end{abstract}

\begin{IEEEkeywords}
	WLAN discovery, heterogeneous network, channel scan, power consumption.
\end{IEEEkeywords}


%
\IEEEpeerreviewmaketitle

\section{Introduction}
The Third Generation Partnership Project (3GPP) Long Term Evolution (LTE) in the recent past has extensively laid its focus on the inter-working
with other Radio Access Technologies (RATs), owing to limited spectrum availability
and exponentially increasing data traffic. The operation in the unlicensed spectrum has specially attracted great
interest, leading to the development of licensed-assisted access to unlicensed spectrum \cite{3gpp_36889} and 3GPP System to Wireless Local
Area Network (WLAN) Inter-working \cite{3gpp_23234}. In 3GPP/WLAN inter-working, a user associated with a 3GPP LTE network can be allowed
to steer/offload its data traffic through a selected IEEE 802.11 \cite{IEEE} based WLAN.
This solution has been referred to as a mechanism to increase the overall capacity of the network, for small capital expenditure.
Further, since most of the commercially available User Equipments (UEs) already support IEEE 802.11 
compatible WLAN interfaces, the transition towards
the inter-working technology, in the context of the user, may only require software up-gradation. Hence, this proposal is widely popular
with 3GPP network operators and UE manufacturers alike.

Despite well defined specifications of IEEE 802.11 WLAN and 3GPP LTE, the optimal resource utilization for
the inter-working scenario still requires tight integration in the context of cross-network information provision and parameter estimation.
While, current research efforts towards achieving this objective are mainly focused on throughput maximization \cite{arghya}
and traffic steering policies \cite{Shashi}, very little emphasis has been laid on the network discovery aspects of the
inter-working. In this paper, we specifically concentrate on the energy efficient discovery of the IEEE 802.11 WLANs.

The discovery process of WLANs comprises of searching suitable WLANs, which satisfy certain Quality of Service (QoS)
criteria, by scanning a subset of the frequency channels supported by the UE's WLAN interface.
Existing WLAN discovery scheme, implemented in most of the commercially available UEs, involves 
scanning all the supported channels of the WLAN interface at regular intervals of time. Each scan operation may consist of scanning
11 channels consecutively for a UE operating in 2.4 GHz band and greater than 20 channels for a UE 
supporting both 2.4 GHz and 5 GHz band. The power consumption for the channel scan may range from 300 to 1400 mW, based
on the UE characteristics and the mode of scanning \cite{Budzisz}.
Since, increasing the number of channels to be scanned leads to higher power consumption of the discovery scheme, the battery life
of the UE may suffer as a consequence. Studies, \cite{Footprint}, have shown that the existing discovery scheme may contribute upto 20\% of
the overall power consumption of a UE. As, high fraction of UEs today support both 2.4 GHz and 5 GHz frequency band,
there is a need to
reduce the number of channel scans to improve the overall user experience.

Since, IEEE 802.11 based WLANs operate in the unlicensed spectrum, the operator deployed WLANs co-exist with the privately deployed WLANs.
Moreover, the number of privately deployed WLANs is expected to outnumber the operator
deployed WLANs, due to high penetration of IEEE 802.11 WLANs.
Hence, to mitigate co-channel interference, the operator deployed WLANs may occupy a small fraction of frequency channels in the unlicensed
spectrum for each location.
Therefore, for such a case, the discovery schemes implemented in commercial UEs result in sub-optimal performance and require further optimization in the context
of scan scheduling and channel selection. We aim to reduce the number of channel scans without compromising the
WLAN offload opportunities for the user.

In this paper, we have proposed an energy efficient IEEE 802.11 WLAN discovery scheme. 
Further, we have performed extensive field experiments on Android platform, along with simulation and analytical studies, to analyze the performance of our 
discovery scheme. Using the experimental results, we have shown the performance gains of our proposed scheme in comparison to other solutions in the literature.

\subsection{Related Work}
Different solutions have been proposed to address the problem of designing efficient neighbor discovery schemes in the context of sensor and mobile wireless networks. Existing neighbor discovery schemes can be divided into two categories, namely probabilistic and deterministic protocols. The basic idea behind probabilistic schemes, \cite{Khalili, McGlynn}, is letting each node in a network transmit, receive or sleep with a certain probability in each slot.
In deterministic procedures \cite{Jiang, Sun}, a fixed active-sleep schedule is involved for each node. 

Other solutions proposed in the literature can be classified based on the optimization objective undertaken.
One set of proposals targets the optimization of the time interval between successive
scan operations, using different criteria. The authors, in \cite{Youngkyu} and \cite{Kim}, determine the interval based on the user speed and
the WLAN spatial density. The scan operations are performed more frequently as the WLAN spatial density increases.
The scheme is expected to reduce the power consumption for small values of WLAN spatial density, but may not
provide substantial benefits as the density increases.
Other solutions, \cite{Guo,Thota}, rely on position and mobility information of the UE, for optimal scan scheduling.
The underlying approach of these solutions is to trigger the scan operation whenever the scheme detects UE motion
or WLAN proximity. While, these schemes perform very few scan operations for a static UE, the power consumption
is likely to increase rapidly as the speed of the UE increases.

The second set of solutions \cite{Castignani,Chang,Waters} attempts to reduce the time spent in the active state of WLAN interface of a UE during a channel 
scan, thereby, decreasing the power consumption.
The authors, in \cite{Castignani}, determine minimum values for MinChannelTime and MaxChannelTime, without affecting the
discovery performance. 
The scheme, in \cite{Waters}, optimizes the number of probe request transmissions for
each channel, based on the previous channel activity.
Although, these solutions lead to lower power consumption, they do not address the issue of energy wastage
due to channel scans which do not provide potential offload opportunities for the UE.

The third set of proposals \cite{Footprint,Shin,Liu,Doppler,Lim} targets the number of channels scanned during each scan operation. 
The scheme in \cite{Shin}
places more emphasis on the non-overlapping frequency channels and scans selective overlapping channels.
The methodology of preferring non-overlapping channels, however, may not
provide sufficient gain for the 5 GHz band, where all channels are non-overlapping. 
A 3GPP network
assisted scheme is proposed in \cite{Doppler}, where, the UE selectively scans the operating channels of WLANs deployed within the coverage
of current 3GPP serving base station,
while, the scheme, described in \cite{Lim}, also
takes into account the position information of the UE and the deployed WLANs,
for efficient channel selection. 
While, the given solutions have shown significant performance improvements, none of the works has been able to validate
the results using actual UE measurements.

The solutions described above, require mobility information of the UE. Some works, \cite{Guo, Lim}, \
have explicitly used the Geo-spatial Positioning System (GPS) module of the UE, for mobility estimation.
While, the GPS readings may provide accurate mobility information, the GPS module operation itself is a power intensive task. 
Other works\cite{Footprint,Doppler,3gpp_36331,Mahima} explore more energy aware sources for mobility estimation.
In \cite{Doppler}, the mobility of the UE is estimated by using accelerometer readings. 
3GPP specifications, \cite{3gpp_36331}, describe a mobility state estimation scheme which is
further enhanced in \cite{Mahima}, to address the issues arising from varying cell sizes and handover failures. But, the solution
only provides three states of mobility, which is not pertinent in many scenarios.

UE position estimation solutions \cite{Kos,Otsason} based on 3GPP network signal strength require
accurate knowledge of path loss characteristics and extensive training data set for each location.
The authors, in \cite{Footprint}, have proposed relatively simple algorithm
for mobility estimation. The algorithm uses the count of number of base stations for which signal strength crosses
a specific threshold. The authors address the issue of signal strength fluctuations by
computing the mean and variance of signal strength for each location, which is then used for mobility estimation. The solution
assumes availability of significant data for variance computation at each location, which is not possible for a high speed user.
Hence, an enhanced solution is required for mobility estimation of the UE, which also takes into account the signal strength fluctuations
observed by a mobile UE.
\subsection{Our Contributions}
In this paper, we have proposed an energy efficient IEEE 802.11 WLAN discovery scheme, 
which uses 3GPP network assistance along with
the results of past channel scans, to optimally select the next channels to scan.
We have also developed a mobility estimation procedure,
based on the 3GPP network signal strength variation patterns, to accurately detect the UE's mobility using a short decision interval. 
Although, we have used the proposed mobility estimation procedure for WLAN scanning, any other mobility estimation procedure from the literature 
can be used for the purpose. 
Note that, the focus is more on energy efficient discovery of IEEE 802.11 WLANs rather than on mobility estimation procedure. 

We have implemented various discovery schemes in Android framework, for accurate performance comparison of our proposed scheme 
against other solutions in the literature. Since, Android does not support selective scanning
mode, i.e, scanning a subset of frequency channels supported by the UE's WLAN interface, we have also developed modules for 
selective scanning support in Android. We have used Google Nexus 5 cell phone as an example but our algorithm can be implemented on any Android 
phone and is expected to produce a similar performance. The power 
consumption is monitored using an Android profiler tool.
Further, we have used different mobility and WLAN deployment scenarios to investigate the performance gains of our proposed scheme.
Note that, while most works in the literature have demonstrated the performance of their schemes using simulations
\cite{Wcnc,Doppler,Lim},
no published literature, to the best of our knowledge, has characterized the power consumption and discovery performance of a 
real UE for the case of a mobile user.

The rest of the paper is organized as follows. 
Section II gives an overview of the IEEE 802.11 based scanning procedure
and 3GPP network assisted offloading. Section III discuses the proposed discovery scheme.
Section IV contains the system model and the experimentation results. Finally, Section V concludes the paper.

\section{WLAN Scanning and Offload Essentials}
\subsection{IEEE 802.11 WLAN Scanning Procedure}
UE's WLAN interface, whether in disassociated or associated state, scans all the supported IEEE 802.11 frequency channels
at regular intervals of time, called the scan interval. Android 5.0 sets the default value of scan interval as
15 seconds for disassociated state and 20 seconds for associated state of the WLAN interface. While, small
values of scan interval results in faster discovery of WLANs, it may significantly reduce the battery life of the UE.
Each channel scan can be carried out in active or passive mode \cite{IEEE}. In passive
mode, the UE keeps its WLAN interface in \emph{Active Listen} state, to receive the beacon frames transmitted
by the WLAN Access Points (WAPs). Since, the time interval between two successive beacon frames from a given WAP is around 100 ms,
the UE has to keep its WLAN interface in \emph{Active Listen} state for a minimum of this duration in order to receive beacons
from all WAPs within its transmission range.
The active scanning mode involves the UE broadcasting probe requests and keeping the WLAN interface in \emph{Active Listen}
state for a minimum time duration, denoted by MinChannelTime, to receive at least one response from a nearby
WAP. The UE extends the duration to MaxChannelTime, if atleast one probe response is received by the
WLAN interface. This is done to ensure that the UE may receive probe responses from other WAPs, which may have been delayed due to
congestion in the network.
The active scanning mode is significantly faster compared to the passive mode, because of lower values of MinChannelTime and MaxChannelTime.
This leads to higher power consumption values for passive scanning mode.

%
\subsection{3GPP Assisted Vertical Handover Procedure}
3GPP, in \cite{141846,141705}, has specified the policies for traffic steering/offload procedure
between a 3GPP network and IEEE 802.11 based WLAN. The policies defined are mainly focused on the signal strength and traffic load
of the respective networks. These are indicated by parameters such as Received Signal Reference Power (RSRP)
of 3GPP serving base station and Received Signal Strength Indicator (RSSI) and channel utilization for IEEE 802.11 based
WLANs. We have concentrated our study on WLAN RSSI threshold criteria.
The traffic steering policy considered in this paper assumes that UE is eligible for initiating offload procedure, if,
the observed RSSI value for a WAP is higher than $S_{enter}$.

In order to achieve fast discovery of WLANs, 3GPP network provides operational
attributes like operating channel and geo-spatial coordinates of the WAPs deployed by the network operator \cite{3gpp_andsf}.
This information is retrieved by the UE by sending a network query along with its current
position to the Access Network Discovery and Selection Function (ANDSF) server.
The position of the UE can be indicated using either serving 3GPP base station identification code or
by stating explicit geo-spatial coordinates of the UE. The UE can also provide identification of a detected WAP,
deployed by the 3GPP network operator, as a position reference.
%
%

\section{WLAN-aware Discovery Scheme}
\begin{figure}[ht]
	\centering
	\resizebox{3.75cm}{!}{
	\begin{tikzpicture}[node distance = 1.5cm, auto,font=\scriptsize, outer sep=0]
		\node [cloud] (disAssoc) {Disassociated};
		\node [block, below of=disAssoc] (initTimer) {Start Scan Timer};
		\node [block, below of=initTimer] (posEst) {\emph{Position Estimation}};
		\node [block, below of=posEst, node distance=1.5cm] (initSelect) {\emph{Initial Channel Selection}};
		\node [block, below of=initSelect, node distance=1.5cm] (channelElim) {\emph{Channel Elimination}};
		\node [block, below of=channelElim, node distance=1.5cm] (scanOper) {Scan Selected Channels};
		\node [decision, below of=scanOper, node distance=2cm] (rssiCrit) {Check RSSI Criteria \\for Association};
		\node [block, below of=rssiCrit, node distance=2 cm] (sleepMode) {Wait for Scan Timer Expiry};
		\node [cloud, below of=sleepMode, node distance=1.5 cm] (Assoc) {Associated};

		\path [line] (disAssoc) -- (initTimer);
		\path [line] (initTimer) -- (posEst);
		\path [line] (posEst) -- node {UE Position} (initSelect);
		\path [line] (initSelect) -- node {InitSelect} (channelElim);
		\path [line] (channelElim) -- node {ScanSet} (scanOper);
		\path [line] (scanOper) -- node{Scan Results} (rssiCrit);
		\path [line] (rssiCrit) -- node [near start] {no}(sleepMode);
		\path [line] (rssiCrit) -- node [near start] {yes} +(-2,0) |- (Assoc);
		\path [line] (sleepMode) -- node {} +(2.5,0) |- (initTimer);
	\end{tikzpicture}
	}
	\caption{WLAN-aware discovery scheme flowchart.}
	\label{scan_flowchart}
\end{figure}
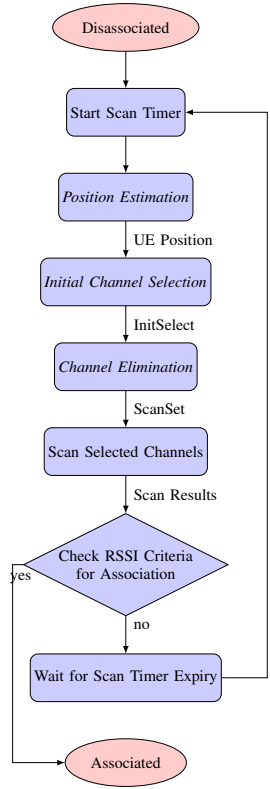

The proposed solution, WLAN-aware discovery scheme, comprises of three steps: Position Estimation,
Initial Channel Selection and Channel Elimination. The entire flow of the scheme is
depicted in Fig \ref{scan_flowchart}. 
The UE, in disassociated state of the WLAN interface, initiates the process
by setting a scan timer. The timer is used to trigger scanning at regular intervals of time.
After setting the scan timer, the UE executes Position Estimation to determine the geo-spatial
coordinates of the UE. It also estimates the distance traveled by
the UE between successive scans, using 3GPP network signal strength measurements, which is used in Initial Channel
Selection and Channel Elimination.
Initial Channel Selection, then, determines the WAPs that lie within a certain proximity of the UE, 
and constructs a list of
operating channels ($InitSelect$) of the selected WAPs.
The proximity of a WAP is determined by computing the distance between the UE's current position estimate and the position of
the given WAP.
Channel Elimination prepares a list of frequency channels ($ScanSet$),
which is a subset of $InitSelect$.
The elimination criteria is based on the WLAN offload policy, and exploits the results of
previous scan operations.

A single scan operation comprises of scanning the channels included in $ScanSet$ by changing the WLAN interface
to \emph{Active} state.
After scanning the channels in $ScanSet$, 
if the UE is able to discover a WAP satisfying the offload RSSI criteria, the UE triggers the association procedure with
the selected WAP, otherwise, it changes its WLAN interface to \emph{Sleep} state. The UE does not perform any further scanning
until the scan timer expires, after which the UE re-initiates the same procedure.

\subsection{Position Estimation}
Position Estimation comprises of determining the geo-spatial coordinates of the UE.
We use the results of the previous scan operations to determine UE's geo-spatial coordinates and hence, the
procedure has negligible impact on the power consumption of the discovery scheme.
We also monitor the distance traveled by the UE between successive
scan operations, which is used to indicate the UE's position error due to mobility.

\subsubsection{Mobility Estimation}
Let $n^{th}$ instance of the scan operation be initiated at time $t_n$.
To estimate the mobility state of the UE between $(n-1)^{th}$ and $n^{th}$ scan operation,
we compute the mean ($\bar{s}_n^i$) and standard deviation ($\sigma_n^i$) of the signal strength
values obtained between time instances $t_{n-1}$ and $t_n$, for each base station $i$.
We only concentrate on classification of two mobility states, i.e.,
a static user ($S_{stat}$) and a mobile user ($S_{mob}$). Since, our focus is on pedestrian mobility scenario, we have determined the mobility estimation parameters of $S_{mob}$ state pertaining to a user traveling with a speed of 1 m/s. Hence, all users with speed greater than or equal to 1 m/s are considered as mobile users.
Let $\phi_n^i$ be defined as,
\begin{equation*}
	\phi_n^i = |\bar{s}_n^i - \bar{s}_{n-1}^i|.
\end{equation*}

Since, the values of $\bar{s}_n^i$, $\phi_n^i$ and $\sigma_n^i$ vary with $n$ and mobility state of 
the UE, we can determine their probability distributions. The probability distribution for the given parameters can be determined by collecting extensive 3GPP network signal strength values for each mobility state of the UE. 
Now, using the values of $\phi_n^i$ and $\sigma_n^i$, we determine a metric $\Delta_{n}$, given by,

\begin{equation}\label{mobility_estimation}
	\Delta_{n}=\cfrac{\sum_{i=1}^{N_{bs}}(e^{-\phi_n^i/\phi_0} + e^{-\sigma_n^i/\sigma_0})}{2 N_{bs}} ,
\end{equation}
where, $N_{bs}$ denotes the number of base stations observed
during interval $t_{n-1}$ to $t_n$. The values of $\phi_0$ and $\sigma_0$ are determined using the probability distributions of
$\phi_n^i$ and $\sigma_n^i$ for each mobility state of the UE. Let $\phi_{50}$ and $\sigma_{50}$ correspond to the
values of $\phi_n^i$ and $\sigma_n^i$, respectively, for which the probability of UE being in $S_{stat}$ is 50\%. The values for $\phi_0$ and 
$\sigma_0$ are then computed as
	$\phi_0 = 1.44 * \phi_{50}$ and $\sigma_0 = 1.44 * \sigma_{50}$.
This is done to ensure that the solutions of $e^{-\phi_n^i/\phi_0} = 0.5$ and $e^{-\sigma_n^i/\sigma_0} = 0.5$,
correspond to 50\% probability of UE being in state $S_{stat}$.

In Equation \ref{mobility_estimation}, we have used exponential metric
as it has been found to give good estimates of mobility through exhaustive simulations and field experiments.
Moreover, fading is the predominant cause of signal strength variation between consecutive scan instances for a pedestrian mobility scenario. 
Since, shadowing variation and fast fading components are generally approximated as log normal and Rayleigh distribution, respectively, the probability distribution 
of signal strength variation is likely to follow an 
exponential trend.    

The mobility state of the UE is determined as static ($S_{stat}$) if the value of $\Delta_{n}$ is greater than 0.5, otherwise,
the UE is classified as mobile ($S_{mob}$).
The distance traveled by the UE, between $(n-1)^{th}$ and $n^{th}$ scan operation, is determined as,
\begin{equation*}
	D_n =  \begin{cases}
		(t_n - t_{n-1}),          & \quad \text{if } \Delta_n \leq 0.5,\\
		0,                        & \quad \text{if } \Delta_n > 0.5.\\
	\end{cases}
\end{equation*}

%

\subsubsection{Position Update Procedure}
Let the geo-spatial coordinates of the UE at $(n-1)^{th}$
scan operation be denoted by $Pos_{n-1}$ and its corresponding error term
by $\Delta_{n-1}$.
As described in Algorithm \ref{posEstimation}, the UE selects the WAP with the maximum observed RSSI, among the set of WAPs discovered
in $(n-1)^{th}$ scan operation. The UE, then, estimates its distance from the selected WAP,
denoted by $\Delta_{wap}$, using the RSSI value of the selected WAP and a known path loss model. 
If $\Delta_{wap}$ is less than  $\Delta_{n-1}$,
the UE updates its geo-spatial coordinates
to the position of the selected WAP, denoted by $Pos_{wap}$. The UE also updates the position error term ($\Delta_n$)
to the summation of $\Delta_{wap}$
and $D_n$. Here, $D_n$ represents the distance traveled by the UE between $(n-1)^{th}$ and $n^{th}$ scan operation.
The value of $D_n$ is determined by the Mobility Estimation procedure, as explained above.
If the value of $\Delta_{wap}$ is greater than $\Delta_{n-1}$,
the updated position error term ($\Delta_n$) is set as the summation of
$\Delta_{n-1}$ and $D_n$. The geo-spatial coordinates of the UE is kept unchanged for this case.

\begin{algorithm}[h]
	\caption{Position Update Procedure}
	\label{posEstimation}
	\begin{algorithmic}[1]
		\State $Disc_{n-1}:$ WAPs discovered in $(n-1)^{th}$ scan operation
		\State $wap\gets$ Find WAP with maximum RSSI in $Disc_{n-1}$
		\State $Pos_{wap}\gets$ Geo-spatial coordinates of $wap$
		\State $\Delta_{wap}\gets$ Distance between $wap$ and UE based on RSSI
		\If {$\Delta_{n-1} > \Delta_{wap}$}
		\State {$Pos_n \gets Pos_{wap}$}
		\State {$\Delta_n \gets \Delta_{wap} + D_{n}$}
		\Else
		\State {$Pos_n \gets Pos_{n-1}$}
		\State {$\Delta_n \gets \Delta_{n-1} + D_n $}
		\EndIf
	\end{algorithmic}
\end{algorithm}

If the GPS coordinates of the UE are available, the UE sets its position and error term as the geo-spatial coordinates and error values
provided by the GPS module, respectively.

\subsection{Initial Channel Selection}
In this stage, the UE constructs a list ($InitSelect$) of operating channels of the WAPs deployed within 
the serving 3GPP base station, such that, the UE is within the 
transmission range of the selected WAPs.
If the transmission range of the deployed WAPs is considered to be $D_r$, then, the UE selects those WAPs whose distances
from the current estimated position of the UE are less than $D_{r} + \Delta_{n}$.
The position
error is incorporated to accommodate the scenario, where, the position error component is significantly large due to infrequent position
updates. The position information of the deployed WAPs can be obtained by the UE by querying the ANDSF server.
The procedure for Initial Channel Selection is shown in Algorithm \ref{initialSelection}.
\begin{algorithm}[t]
	\caption{Initial Channel Selection}
	\label{initialSelection}
	\begin{algorithmic}[1]
		\State {$ServSet:$ List of WAPs within serving 3GPP site}
		\For {each $wap$ in $ServSet$}
		\State {$D_{wap}\gets$ Distance between $wap$ and UE}
		\State {$f_{wap}\gets$ Operating channel of $wap$}
		\If {$D_{wap} < D_{r} + \Delta_{n}$}
		\State {Add $f_{wap}$ to $InitSelect$}
		\EndIf
		\EndFor
		\State Return $InitSelect$
	\end{algorithmic}
\end{algorithm}

\subsection{Channel Elimination}

For each frequency channel ($f_w$) supported by the WLAN interface, the UE stores an associated RSSI value ($S[f_w]$),
which is the maximum RSSI observed by the UE in the previous scan instance of the given channel.
The UE also maintains a distance metric ($D[f_w]$), indicating the
distance traveled by the UE from the time the channel $f_w$ was last scanned.
The procedure is initiated by incrementing the value of $D[f_w]$ for each channel by $D_n$, i.e.,
the distance traveled by the UE between $(n-1)^{th}$ and $n^{th}$ scan operation.
\begin{algorithm}[t]
	\caption{Channel Elimination}
	\label{wlanElimination}
	\begin{algorithmic}[1]
		\For {each $f_w$ in supported channels}
		\State {Increment $D[f_w]$ by $D_n$}
		\State {$S[f_w]\gets$ Maximum RSSI in previous scan of $f_w$}
		\EndFor


		\For {each $f$ in $InitSelect$}
		\If {$S[f] \geq S_{high}$}
		\State {Add $f$ to $ScanSet$}
		\ElsIf {$S[f] < S_{high}$ \textbf{and} $D[f] > D_{1}$}
		\State {Add $f$ to $ScanSet$}
		\ElsIf {$S[f] = NaN$ \textbf{and} $D[f] > D_{2}$}
		\State {Add $f$ to $ScanSet$}
		\EndIf
		\EndFor

		\For {each $f$ in $ScanSet$}
		\State {$D[f] \gets 0$}
		\EndFor

		\State Return $ScanSet$
	\end{algorithmic}
\end{algorithm}
Now, for each channel ($f$) included in $InitSelect$,
if $S[f]$ is greater than $S_{high}$, the channel is added to $ScanSet$. Otherwise, if $S[f]$ is less than $S_{high}$,
the channel is included in $ScanSet$, only if the distance traveled by the UE since the last scan of the given
channel ($D[f]$) is greater than
a threshold $D_{1}$. Similarly, if the value of $S[f]$ is not available, i.e., no WAP is detected in the previous scan of the channel $f$,
the channel is added to $ScanSet$,
if its distance metric $D[f]$ is higher than $D_{2}$. Here, $D_{2} > D_{1}$ to ensure that channels, where WLAN activity is detected,
are scanned more frequently. The value of $S_{high}$ is set higher than $S_{enter}$, i.e., offload RSSI criteria, such that the signal fluctuations due to fading
are accounted in the difference. Finally, for the frequency channels present in $ScanSet$,
the value of $D[f]$ is set as zero.
The procedure for Channel Elimination is shown in Algorithm \ref{wlanElimination}.

\section{Experiments and Results}
\subsection{Experimentation Setup}
\subsubsection{Implementing Selective Channel Scanning}
We have used Google Nexus 5 cell phone as the UE. Nexus 5 employs Broadcom bcm4339 WLAN interface.
The choice of Nexus 5 was dictated due to the support
of Android Open Source Project (AOSP) source building for customized scan implementation. We have used custom built image
of Android version 5.0.2 for the UE operation. 

The current version of Android 5.0 Application Programming Interface (API) does not support selective scanning mode, i.e, scanning
a user configured list of frequency channels.
Hence, our primary challenge was to modify the implementations of AOSP source, to support the selective channel scanning mode.
This is a non-trivial task, since, it requires careful consideration of the native operating system procedures for different states
of the WLAN interface.

\begin{figure}[!t]
	\centering
	\includegraphics[width=0.5\textwidth]{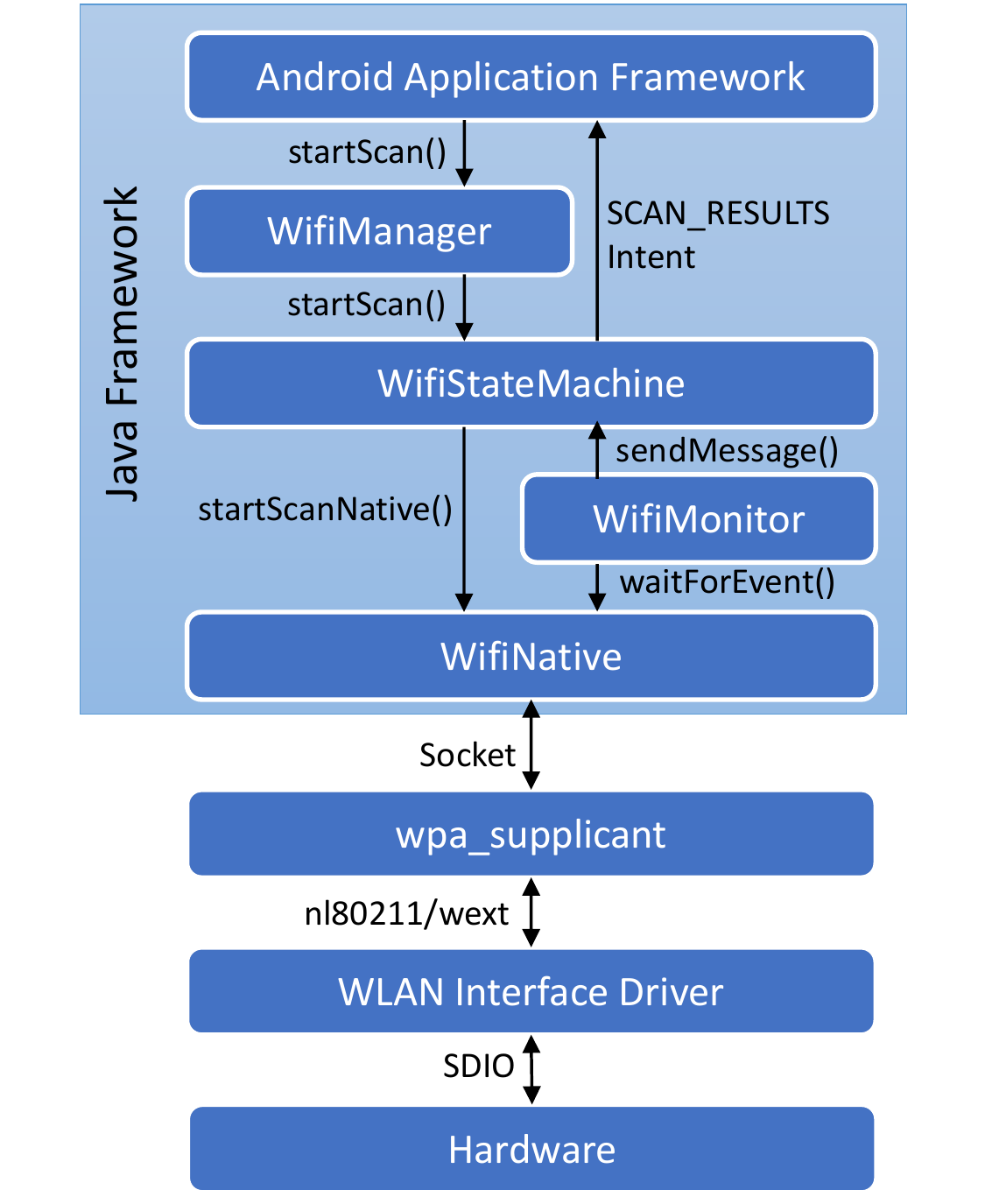}
	\caption{Android WLAN scanning framework.}
	\label{fig_scan_events}
\end{figure}

Fig \ref{fig_scan_events} illustrates the various components which are involved in the WLAN scanning procedure and
the corresponding procedure calls.
As, the discovery schemes may require information from telephony as well as the GPS module, the schemes
are implemented in the Java framework which provides interface to different UE components.
The scan procedure, in the Java framework, is initiated in $WifiManager$ module,
which provides interface to the user for the management of the connections between the UE and the WLANs.
The user invokes $startScan$ routine of $WifiManager$ to initiate the scan operation, which makes successive calls
to $WifiStateMachine$ and $wpaSupplicant$, before invoking the device driver of the WLAN interface.
$WifiStateMachine$ implements the scanning and connection modules for different states of the WLAN 
interface and $wpaSupplicant$ communicates and sends commands to the WLAN interface driver. 

Although, $WifiManager$ does not
support selective scanning mode, $wpaSupplicant$ driver
provides API for scanning user configured channels, which can be called using the Java framework of Android. Hence, we have
implemented modules in $WifiManager$ and $WifiStateMachine$ to incorporate the functionality of selective scanning mode.
Also, as $WifiStateMachine$ implements its own scheduling algorithm for scan operations, we have disabled
the native scan procedure of Android.

An Android device that is left idle quickly falls asleep, to avoid battery drainage, leaving all the running applications
in a standby mode.
User applications use an Android system service called wake lock, to ensure uninterrupted operation.
Wake lock service selectively keeps the CPU on for the time interval it is held by an application.
However, Android's $ActivityManagerService$ terminates any application which holds the wake lock
for a duration greater than a specific threshold. Since, our analysis required elongated background service, we
have modified the $ActivityManagerService$ implementation to disable the process termination routine for the application context
of the discovery schemes.

\subsubsection{Power Consumption Profiling}
While, the inbuilt fuel
gauge of Google Nexus 5 provides accurate readings when determining
battery state-of-charge over a long period of time, i.e., greater than 5 hrs, the error produced due to low resolution
of fuel gauge readings is not
acceptable for shorter time span of operation.
Hence, Qualcomm Trepn Analyzer \cite{Trepn}, which is a power monitoring/profiler tool,
is used to analyze the power consumption of the implemented discovery
schemes. The power consumption readings from the profiler are decremented by the power consumption values of the baseline measurement, where,
only Trepn Analyzer executes in the background with no other activity in running state. Trepn Analyzer is calibrated for Nexus 5 power
consumption and therefore, can provide
useful insights on the power savings of discovery schemes.

\begin{figure}[!t]
	\centering
	\includegraphics[width=0.45\textwidth]{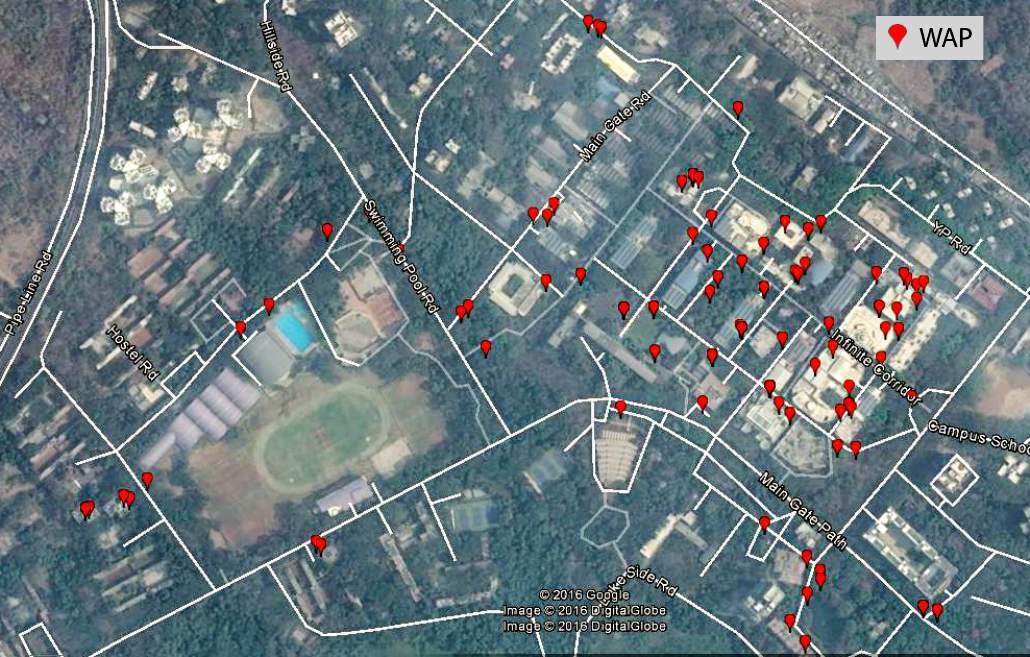}
	\caption{WAP deployment.}
	\label{fig_map}
\end{figure}
\begin{figure}[!t]
	\centering
	\includegraphics[width=0.45\textwidth]{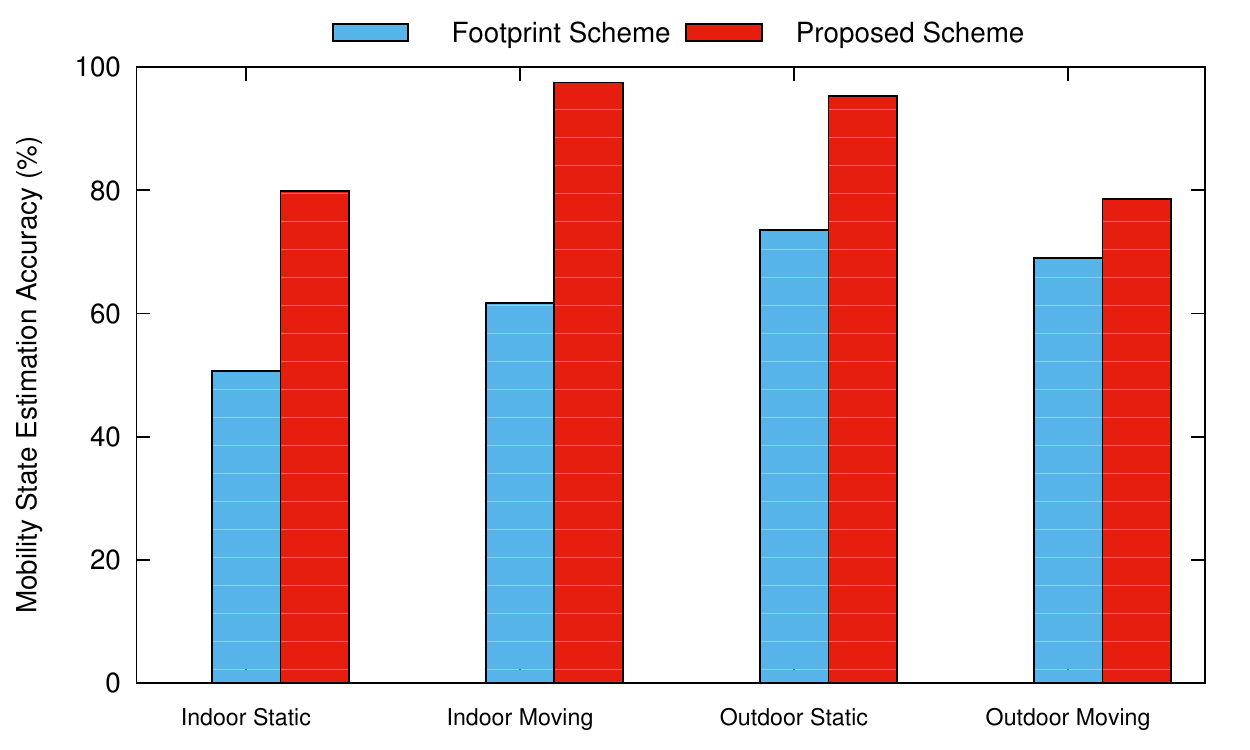}
	\caption{Mobility State Estimation accuracy for different schemes.}
	\label{fig_mobility}
\end{figure}
\begin{figure*}[!tb]
    \centering
    \begin{subfigure}[t]{0.40\textwidth}
        \centering
        \includegraphics[width=\textwidth]{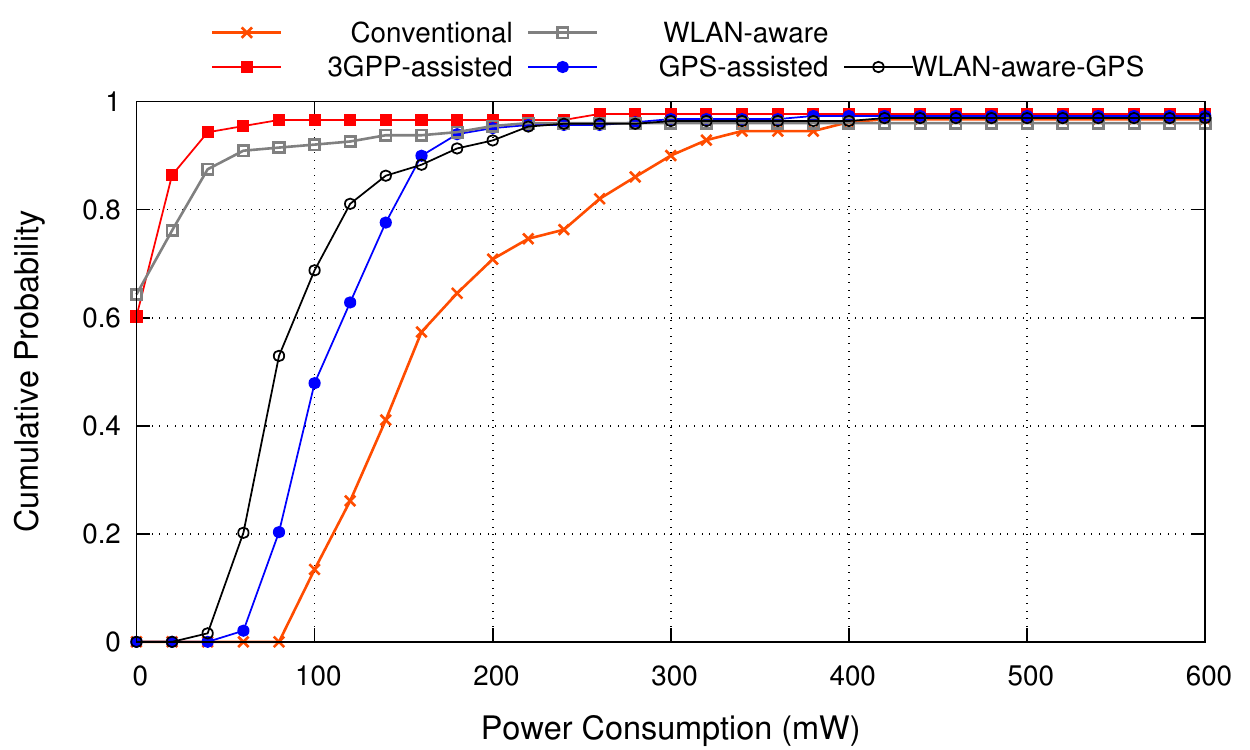}
        \caption{Number of WAPs deployed - 10.}
        \label{fig:power_100_1}
        \end{subfigure}%
    ~ 
    \begin{subfigure}[t]{0.40\textwidth}
        \centering
        \includegraphics[width=\textwidth]{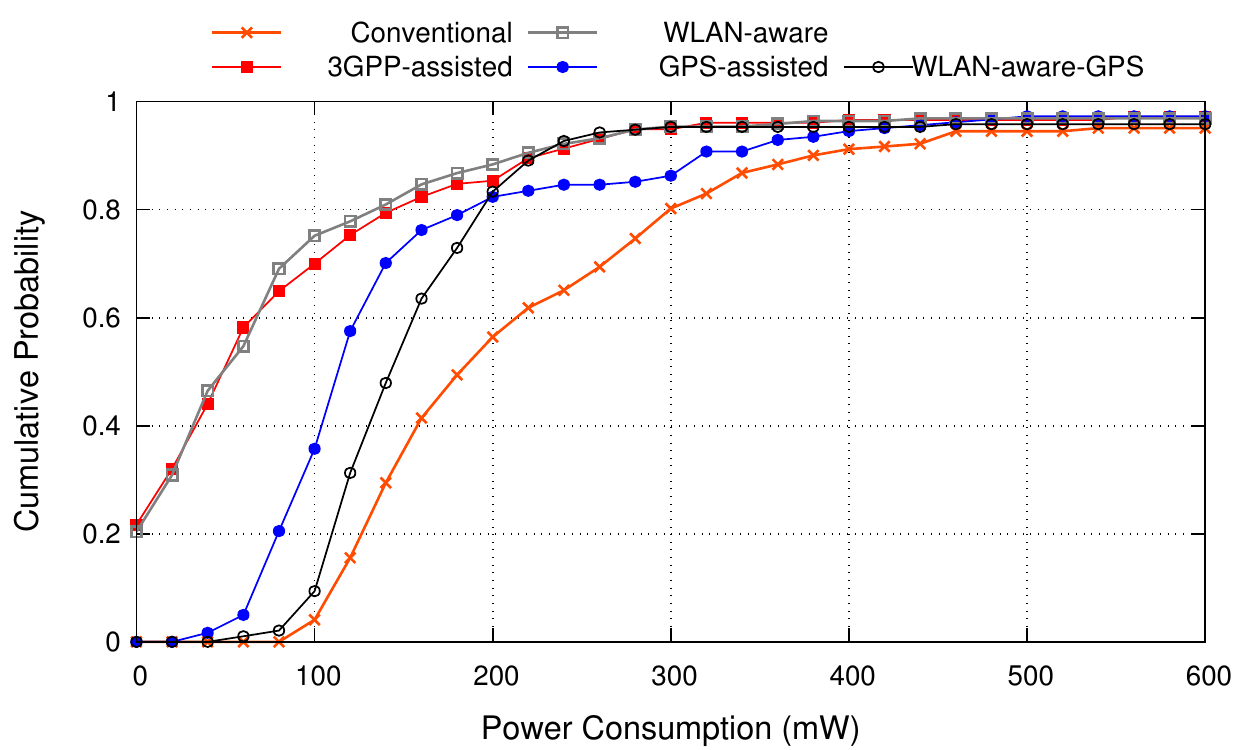}
        \caption{Number of WAPs deployed - 20.}
        \label{fig:power_20_1}
        \end{subfigure}
        \\
    \begin{subfigure}[t]{0.40\textwidth}
        \centering
        \includegraphics[width=\textwidth]{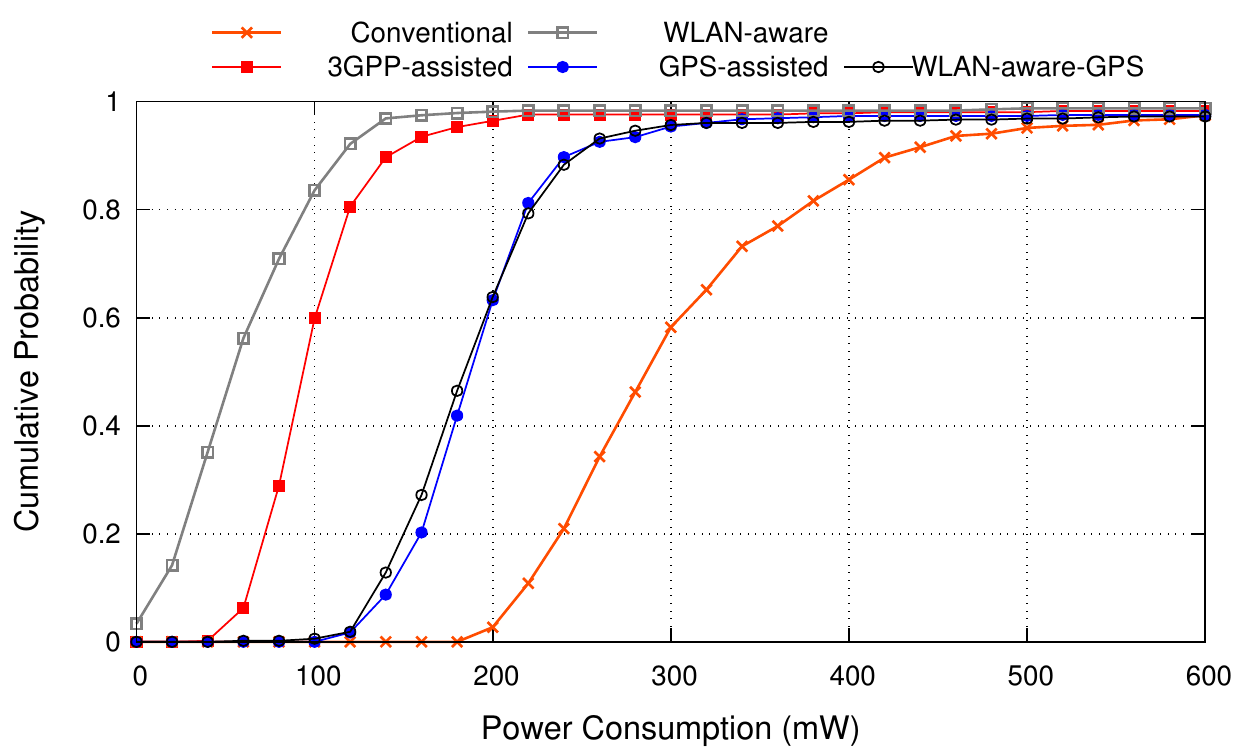}
         \caption{Number of WAPs deployed - 50.}
         \label{fig:power_100_2}
       \end{subfigure}
       ~
       \begin{subfigure}[t]{0.40\textwidth}
        \centering
        \includegraphics[width=\textwidth]{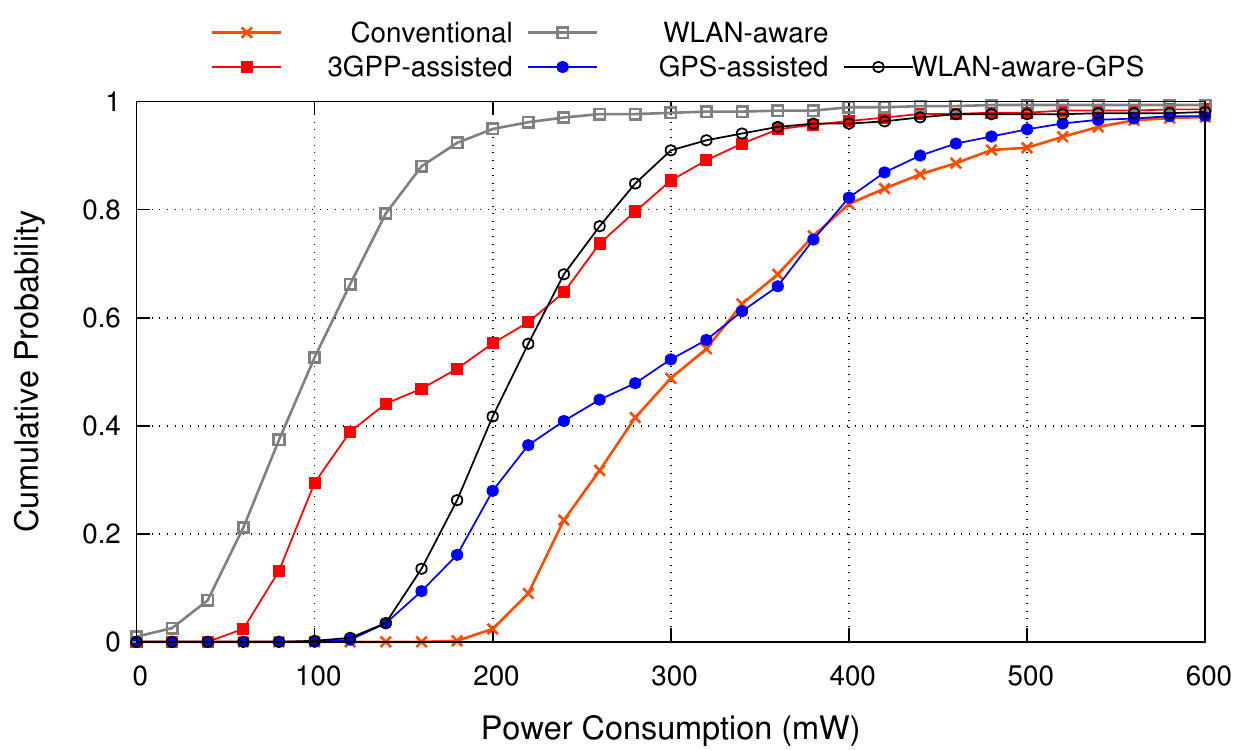}
        \caption{Number of WAPs deployed - 100.}
        \label{fig:power_20_2}
        \end{subfigure}
        \caption{Cumulative probability distribution of UE power consumption in mW, for different discovery schemes and WAP density scenarios. The power consumption values are obtained for each scan interval.}
	\label{fig_power}
\end{figure*}
%
%

\subsubsection{WAP Deployment Topology}
IIT Bombay campus wide WAP deployment has been used as an experimental testbed, for discovery analysis. 
We have selected 100 WAPs for our assessment.
Since, most of the WAPs are deployed inside the campus buildings, we have selected those WAPs such that an outdoor UE experiences high RSSI
from them.
Fig \ref{fig_map} illustrates the WAP deployment topology within the campus. 
For field experiments, we have considered four scenarios where from the 100 WAPs, 
10, 20, 50 and all the 100 WAPs, respectively, are randomly selected.

\subsection{Mobility State Estimation Results}
We have performed field experiments for different mobility scenarios, i.e. indoor moving user, indoor static user, outdoor moving user and outdoor static user, and compared the results of our proposed mobility estimation scheme against the scheme defined in \cite{Footprint}, which we call as Footprint scheme. For a moving user experimentation setup, the user moves with an approximate speed of 1 m/s continuously for a minimum duration of 1 minute per iteration. Mobility state estimation results for various such iterations are compiled to compute average statistics. Similar procedure is followed for the static user. Fig \ref{fig_mobility} shows the mobility state estimation accuracy for different schemes and mobility scenarios. The proposed mobility state estimation scheme outperforms the Footprint scheme in all the mobility scenarios.

Note that, the accuracy of both the mobility schemes is not as good for the Indoor Static and Outdoor Moving scenarios compared to the other two scenarios. For Indoor Static scenario, the user experiences significant signal strength variation even when static, due to penetration losses and pronounced effect of multi-path fading. Hence, this fluctuation may result in decreased accuracy of mobility state estimation. While, for Outdoor Moving scenario, the signal strength variation is not as high as compared to Indoor Moving scenario, which results in user being estimated as static in a number of cases.

\subsection{WLAN Scanning Results and Observations}
We have used test cases with different user mobility routes to analyze the performance of different discovery 
schemes for continuous user mobility. The analysis is performed for the results 
obtained from the compilation of all the test cases.
We have analyzed the performance of the discovery schemes using the previously described WAP deployment scenarios.
Table \ref{final_param} illustrates the parameter values used in the field experiments.
For field experiments, the user travels with the speed of 1 m/s.
Each observation set contains results obtained from field experiments conducted for a duration of 2 hrs along different mobility routes.
\begin{table}[!t]
	\renewcommand{\arraystretch}{1.3}
	\caption{Field Experiment Parameters.}
	\label{final_param}
	\centering
	\begin{tabular}{|p{4em}|p{19em}|p{3.8em}|}
		\hline
		\textbf{Parameter}      &\textbf{Description}                           & \textbf{Value}        \\ \hline
		$v$                     & User Speed                                    & 1 m/s                 \\ \hline
		$T_{s}$                 & Scan Interval                                 & 10 sec                \\ \hline
		$S_{enter}$             & Offload RSSI criteria                         & -75 dBm               \\ \hline
		$D_{r}$                 & Threshold for Initial Channel Selection              & 100 m                 \\ \hline
		$S_{high}$              & RSSI Threshold for Channel Elimination        & -85 dBm               \\ \hline
		$D_{1}, D_{2}$          & Distance Thresholds for Channel Elimination   & 10, 20 m              \\ \hline
	\end{tabular}
\end{table}
We have analyzed the following discovery schemes:
\begin{enumerate}
	\item Conventional Scheme: UE scans all the channels supported by the WLAN interface, in each scan operation.
	\item 3GPP-assisted Scheme: UE scans only the operating channels of the WAPs deployed within the coverage of 3GPP serving
		base station, in each scan operation. The channel information is provided by the 3GPP network itself.
	\item GPS-assisted Scheme: UE scans only the operating channels of the WAPs deployed within $D_r$ distance from
		the current UE's position in each scan operation.
	\item Proposed WLAN-aware Scheme: WLAN-aware discovery scheme without GPS assistance.
	\item Proposed WLAN-aware-GPS Scheme: WLAN-aware discovery scheme with the GPS co-ordinates of the UE
		available for position estimation.
\end{enumerate}
Here, WLAN-aware and WLAN-aware-GPS discovery schemes are presented in Section III of this paper. 3GPP-assisted and GPS-assisted schemes
are presented in \cite{Doppler} and \cite{Lim}, respectively.
\begin{figure}[!t]
	\centering
	\includegraphics[width=0.45\textwidth]{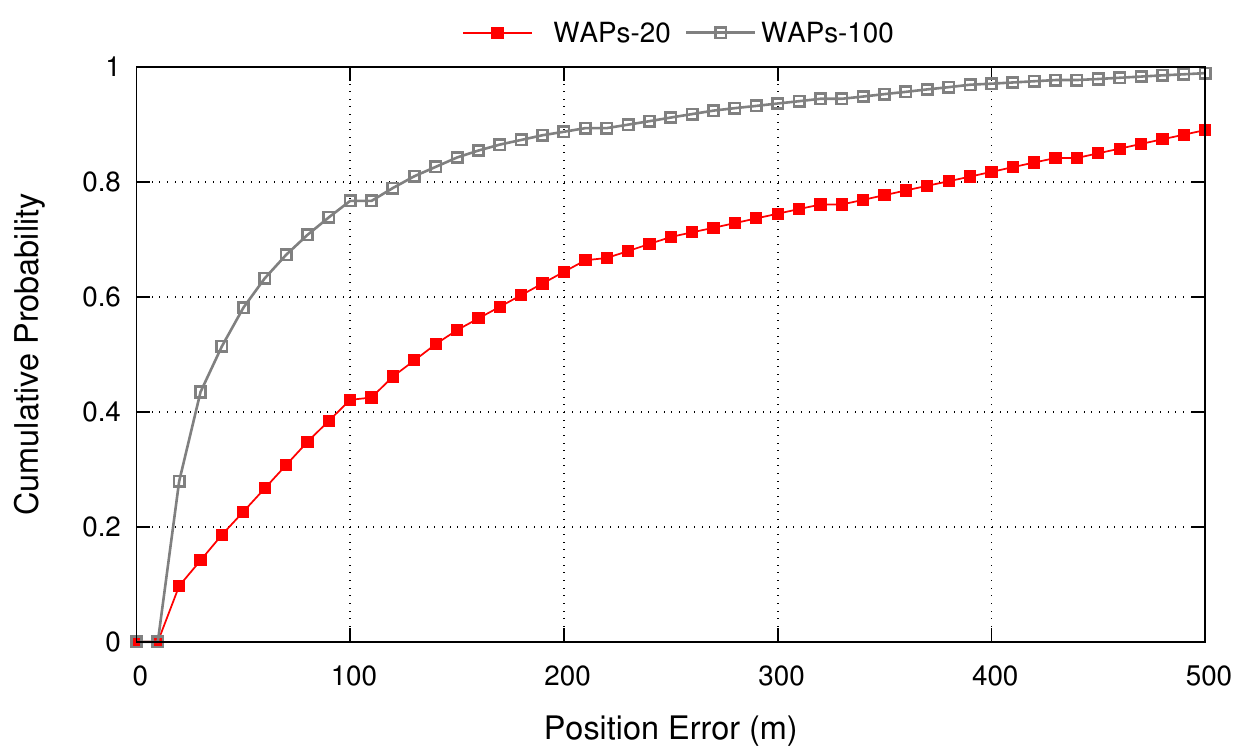}
	\caption{Cumulative probability distribution of position error component of WLAN-aware scheme.}
	\label{fig_position_error_wlan}
\end{figure}
\begin{figure*}[!t]
	\centering
	 \begin{subfigure}[t]{0.45\textwidth}
        \centering
	\includegraphics[width=\textwidth]{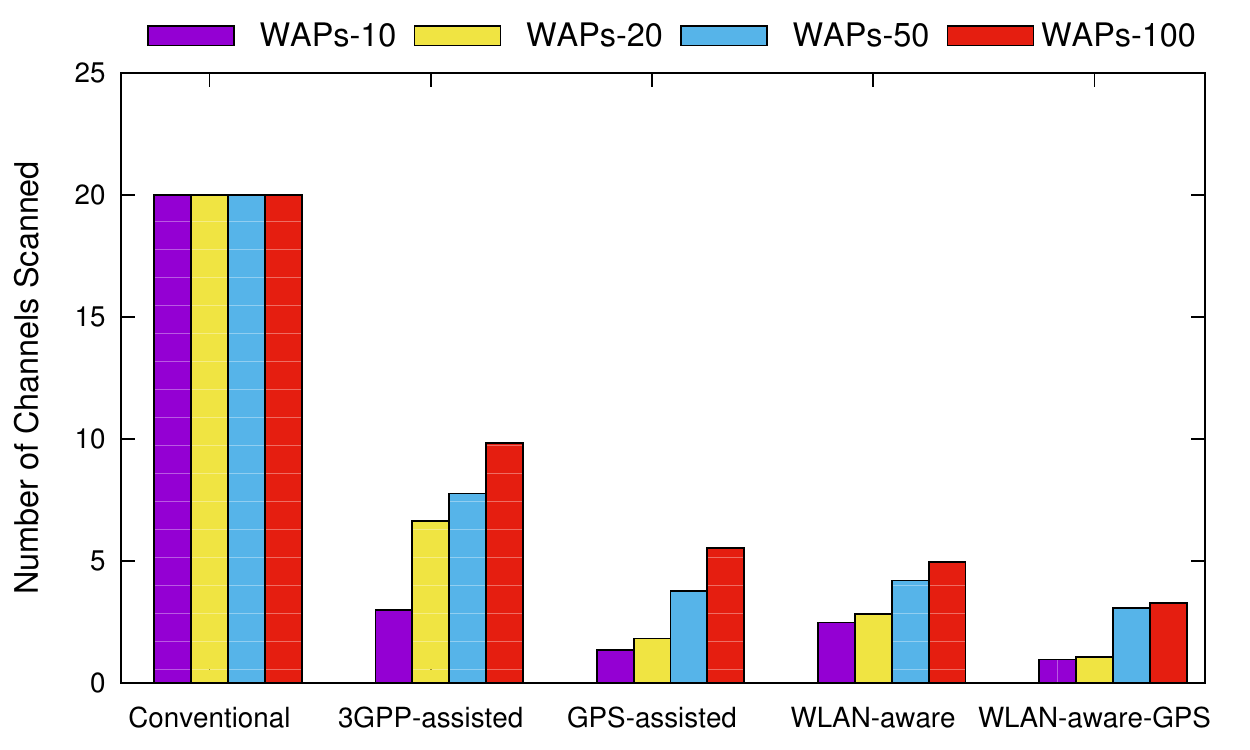}
	\caption{Number of channels scanned per scan operation by different discovery schemes for the various WAP densities.}
	\label{fig_channel_all}
\end{subfigure}
~
 \begin{subfigure}[t]{0.45\textwidth}
        \centering
	\includegraphics[width=\textwidth]{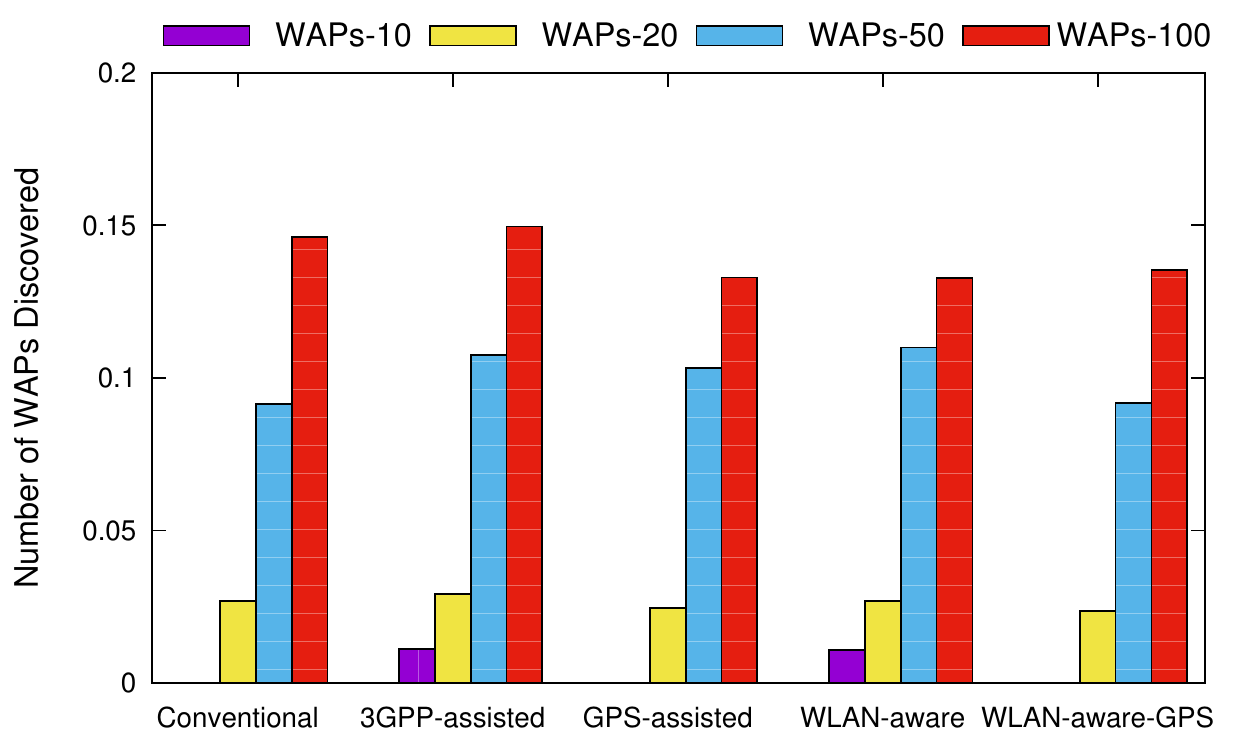}
	\caption{Number of WAPs detected with RSSI $>$ -75 dBm  per scan operation for different discovery schemes.}
	\label{fig_delay_all}
\end{subfigure}
\caption{Discovery performance of different discovery schemes.}
\end{figure*}
\begin{figure}[h]
	\centering
	\includegraphics[width=0.45\textwidth]{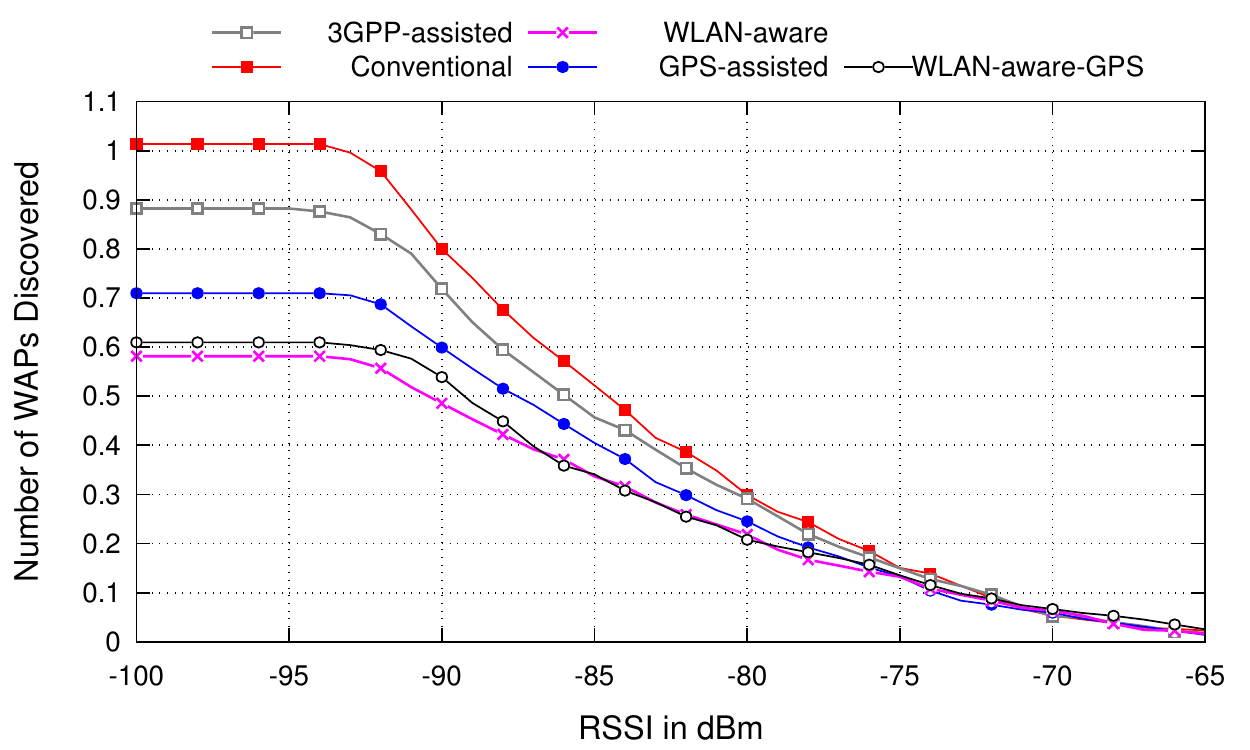}
	\caption{Discovery performance of different discovery schemes for 100 WAPs deployment scenario.}
	\label{fig_channel_wap_100}
\end{figure}

Fig \ref{fig_power} depicts the cumulative probability distributions of the power consumption
values obtained per scan interval for the UE. As evident
from the figures, the Conventional scheme results in highest power consumption, while, WLAN-aware scheme consumes the least power.
The Conventional scheme consumes about 250 mW power for low WAP spatial density, with the value rising to 400 mW as the density increases.
The variation in the power consumption with varying WAP spatial density, even with the same number of channel scans,
is due to different power consumption
of the UE in $Active Listen$ and $Active Receive$ states of the WLAN interface.

The power consumption of the WLAN-aware scheme ranges from 50 mW to 100 mW, as the WAP spatial density increases.
Note that the power consumption value of WLAN-aware scheme is the lowest for the case of 50 and 100 WAPs deployment scenario. This can be explained due to the fact that the UE observes more frequent discoveries,
when the WAP spatial density is high.
Hence, the position error of the UE is quite small. This results in near optimal channel
selection in the \emph{Initial Channel Selection}. On the other hand,
when the WAP spatial density is low, the position error is significantly high and hence, the scheme selects more number of channels for scan. Fig \ref{fig_position_error_wlan} illustrates the position error component of WLAN-aware scheme, determined using Algorithm \ref{posEstimation}, for 20 WAPs and 100 WAPs deployment scenario.

From Fig \ref{fig_power}, it can be observed that the WLAN-aware-GPS and GPS-assisted schemes have identical power
consumption values, for small WAP spatial density. But, the WLAN-aware-GPS performs significantly
better than the GPS-aware scheme, for the scenario of 100 WAPs deployment. 
This can be attributed to the fact that, for small WAP spatial density scenario, the number
of channels scanned by both schemes is significantly low and the main power consuming component is the GPS operation. When
the WAP spatial density is increased, the increase in the number of channels scanned by the GPS-assisted scheme results
in higher power consumption compared to the WLAN-aware-GPS scheme.
Moreover, for the scenario of 100 WAPs deployment, the power consumption of GPS-assisted scheme
is identical to that of conventional scheme, due to high number of channel scans.
Hence, GPS-assisted scheme may not provide sufficient improvement over the Conventional scheme as the WAP spatial density increases.

The 3GPP-assisted scheme has significantly lower power consumption values compared to the GPS-assisted scheme.
But, the power consumption value steadily increases for higher WAP spatial density. The 3GPP-assisted
scheme consumes about 50 mW of power for low values of WAP spatial density, the value increases to about 300 mW for the scenario of
100 WAPs deployment.

Fig \ref{fig_channel_all} shows the number of channels scanned per scan operation by the discovery schemes evaluated.
It can be observed that the WLAN-aware-GPS performs the least number of channel scans compared to other schemes.
The GPS-assisted scheme performs smaller number of channel scans compared to WLAN-aware scheme, for the case of low WAP density scenarios.
But, the WLAN-aware scheme shows significant improvements over the GPS-assisted scheme, for the case of 100 WAPs deployment.
The reason for this is that when the WAP spatial density is low, infrequent WLAN discoveries lead to significant position
error for WLAN-aware scheme and hence, \emph{Initial Channel Selection} takes more number of WAPs in consideration.
But, the position error component is relatively small for higher values of WAP densities, which leads to more
accurate position estimation.
\begin{figure*}[!t]
	\centering
	\begin{subfigure}[t]{0.45\textwidth}
        \centering
	\includegraphics[width=\textwidth]{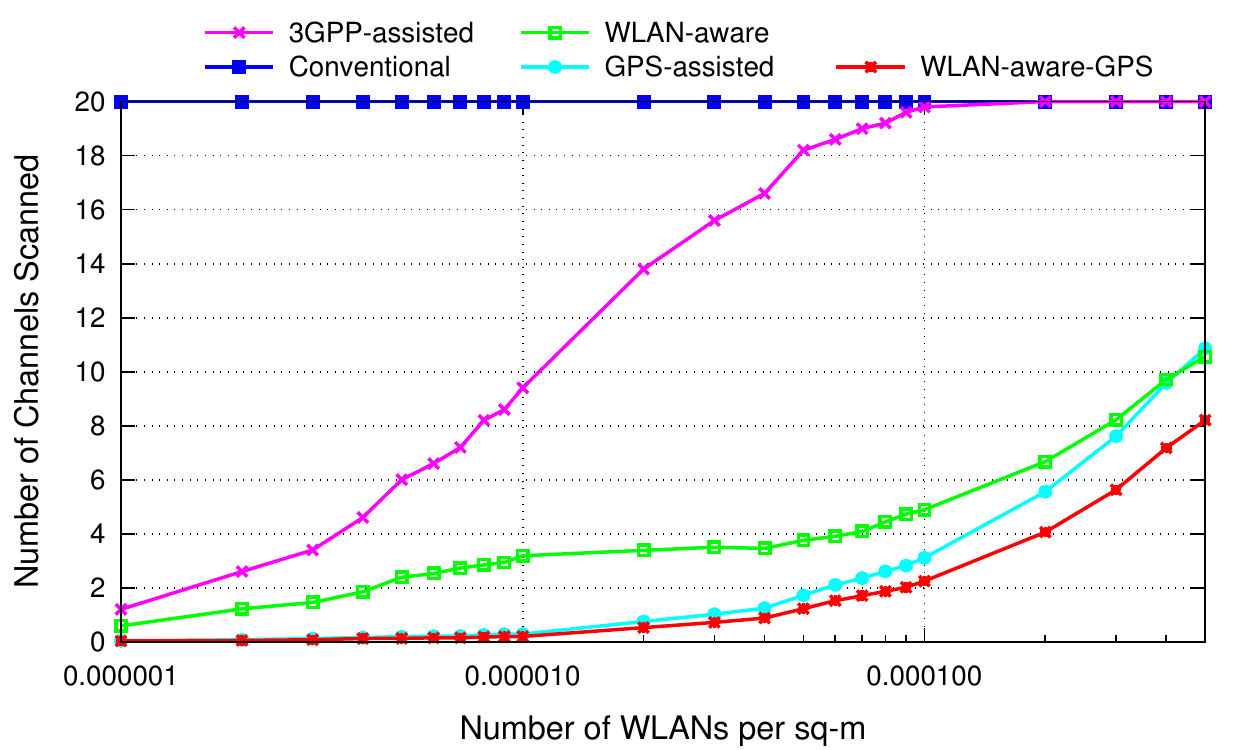}
	\caption{Number of channels scanned per scan operation by different discovery schemes as a function of WAP spatial density.}
	\label{fig_ns_3_channels}
        \end{subfigure}
        ~
        \begin{subfigure}[t]{0.45\textwidth}
	\includegraphics[width=\textwidth]{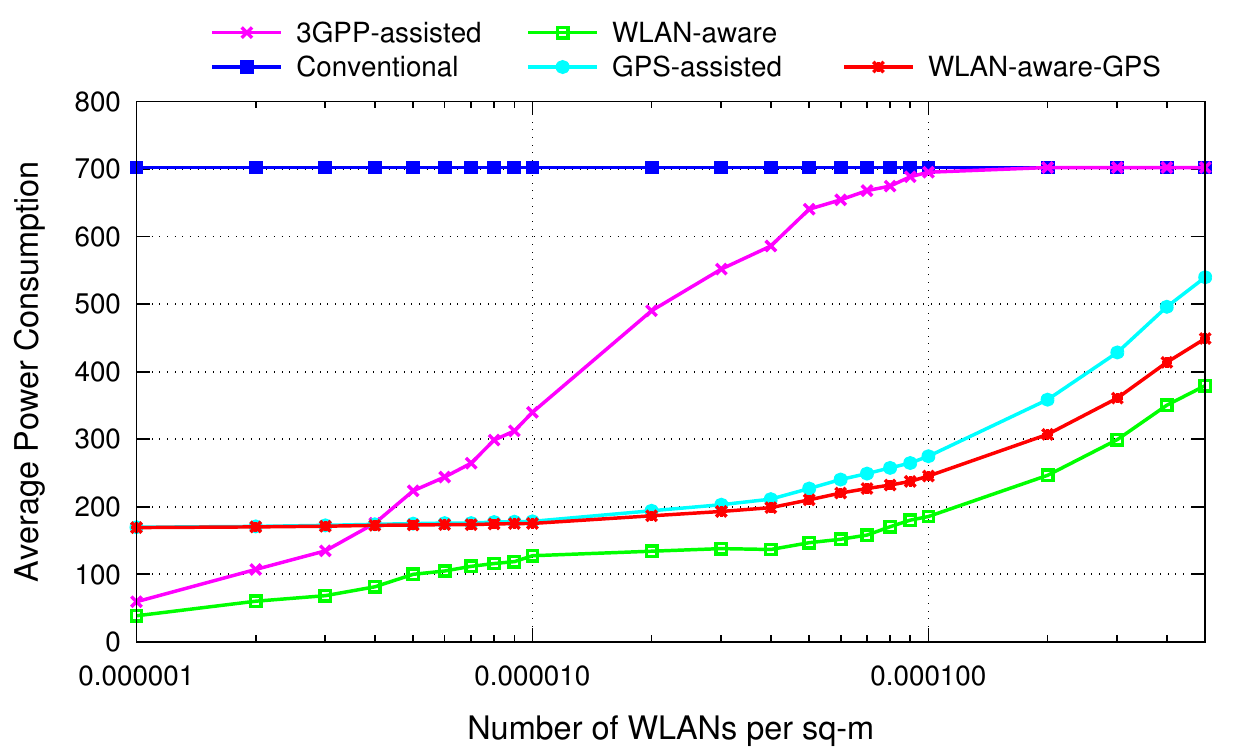}
	\caption{Average power consumption of different discovery schemes as a function of WAP spatial density.}
	\label{fig_ns_3_power}
         \end{subfigure}
         \caption{Discovery performance of different discovery schemes.}
\end{figure*}

Fig \ref{fig_delay_all}, shows the discovery performance of the schemes analyzed. The values on y-axis represent
the number of WAPs detected by the UE per scan operation with signal strength greater than $S_{enter}$.
It can be observed that the discovery performance
of each of the schemes is similar to that of the Conventional scheme. 

Fig \ref{fig_channel_wap_100} depicts the discovery performance for the scenario where 100 WAPs are
deployed in the region. The y-axis indicates the number of WAPs detected by the UE per scan operation, with observed RSSI
greater than the RSSI value indicated by the x-axis. It can be seen that the discovery performance of the
schemes differ substantially, for low RSSI values. But, the difference in the discovery performance diminishes,
for RSSI values greater than -75 dBm.
Since, the 3GPP-assisted scheme scans each operating channel of the WAPs that are deployed within the coverage of the serving
base station, the discovery performance of the 3GPP-assisted scheme is expected to be similar to that of the Conventional scheme.
It can also be observed that the GPS-assisted scheme is not able to detect some of the WAPs (discovered by
the Conventional scheme). This can be explained by noting that
due to inaccurate position information of the WAPs, the UE may not consider 
WAPs for scanning which are around $D_r$ distance from the UE. Moreover, the transmission range of some of the 
WAPs can exceed $D_r$, but, the UE does not consider such WAPs for scan
operation, if their distance is greater than $D_r$.
Since, the UE is positioned near the edge of transmission range of such WAPs, the
UE will experience very low RSSI from the given WAPs. Hence, the GPS-assisted scheme shows loss in discovery performance compared
to the Conventional Scheme.
The WLAN-aware and WLAN-aware-GPS avoid scanning operating channels of the WAPs which are not likely to satisfy 
offload criteria and hence, have lower number of discoveries with observed RSSI less than -75 dBm. 

\subsection{Simulation Analysis}
Although, the conducted field experiments have shown enhanced performance for the WLAN-aware scheme, we have also used simulation study to ascertain the performance gains of the scheme, for wide range of WAP spatial density.
The simulations are performed in Network Simulator 3 (ns-3) which is a link level simulator.

A pedestrian mobility model is considered for simulations, which is based on 800*1200 $m^2$ map of IIT Bombay
campus. The model comprises of a network of intersection nodes and the links between them representing roads on which
a user may navigate. A subset of these intersection nodes are selected as hot spot locations, i.e., points of interest for the
user. At the start of the simulation, the user randomly selects a hot spot node as the destination node and determines the
shortest route to reach there from its current position. The user travels with a constant speed along the selected route and does
not halt at any intermediate intersection node. After reaching the destination node, the user re-initiates the same procedure with
current position of the user becoming the source node.

We have considered a Poisson Point distribution model for WAP deployment.
WAPs operating in each of the channels supported by the UE's WLAN interface are deployed using a Poisson Point distribution with WAP spatial
density, i.e., WAPs per $m^2$, given by $\lambda/n$. The value of $\lambda$ ranges from $10^{-6}$ to $5*10^{-4}$ and
$n$ represents the total number of channels supported by the WLAN interface of the UE. 
As such, the number of WAPs in the entire region ranges from
0 to 500. All the WAPs are assumed to be in the coverage of a single 3GPP base station.
We use the same parameter values as shown in Table \ref{final_param}, for the simulations.
Each simulation set for a given value of $\lambda$ consists of 5 iterations, where in each iteration, a user travels within
the region for a duration of 1 hr.

Fig \ref{fig_ns_3_channels} depicts the number of channels scanned per scan operation by the discovery schemes analyzed
as a function of WAP spatial density.
The number of channels scanned by the 3GPP-assisted scheme rises sharply with the
increasing WAP spatial density, as compared to other discovery schemes.
The number of channels scanned per scan operation saturates to 20 channels,
when the WAP spatial density reaches the value 0.0001 WAPs/m$^2$.
Thereafter, the 3GPP-assisted scheme has identical performance to that of the Conventional scheme.
Also, note that the GPS-assisted and the WLAN-aware-GPS schemes have the least number of channel scans for low values of WAP spatial
density. But, the difference diminishes between the GPS-assisted and the WLAN-aware
scheme, as the WAP spatial density increases. This observation corraborates with the field experimental results discussed in the previous section.
%

Fig \ref{fig_ns_3_power} depicts the power consumption of the discovery schemes, as the WAP spatial density varies. 
The WLAN-aware scheme demonstrates the highest power saving. 
Though, the GPS-assisted and the WLAN-aware-GPS perform the least number of channel scans, the GPS
module operation results in high power consumption for the schemes.
The 3GPP-assisted scheme, for low values of WAP spatial density, has lower power consumption values
in comparison to the GPS-assisted and the WLAN-aware-GPS schemes. But,
as the spatial density of the WAP steadily increases, the higher number of channels scanned by the 3GPP-assisted scheme 
leads to high power consumption values.
\begin{figure}[t]
	\centering
	\includegraphics[width=0.45\textwidth]{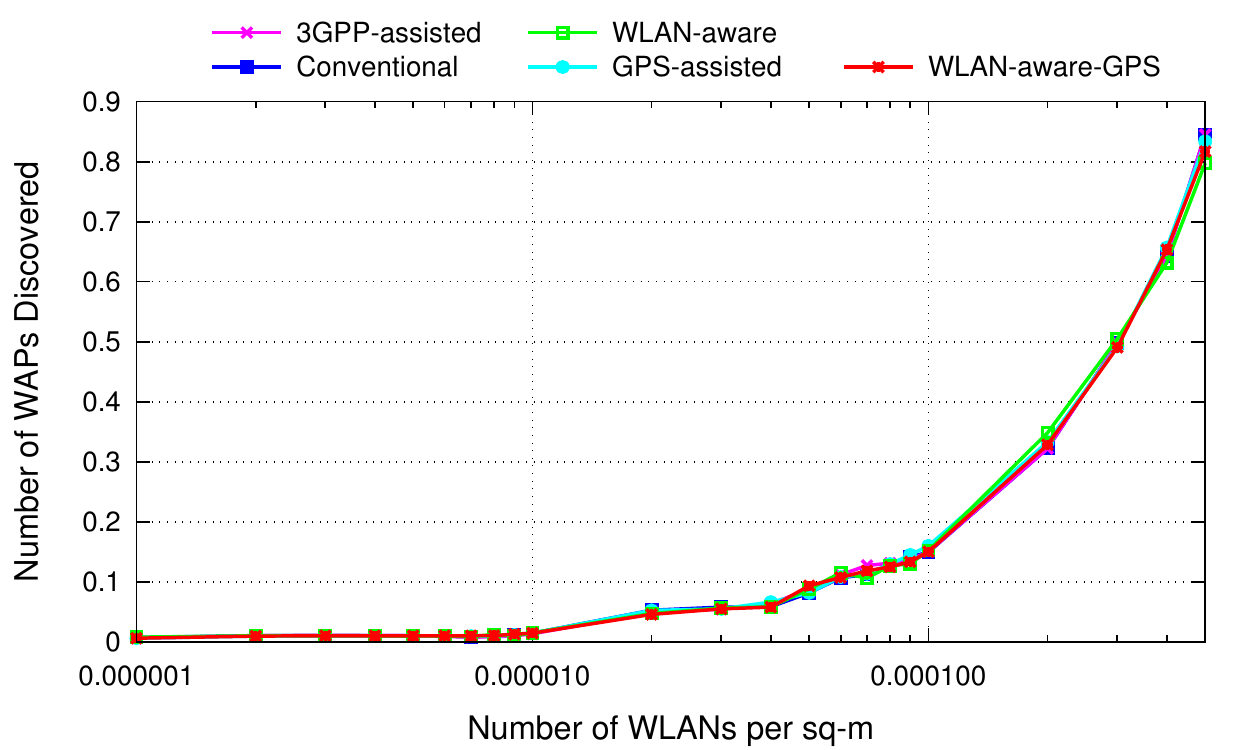}
	\caption{Number of WAPs discovered (with RSSI $>S_{enter}$) per scan operation
	vs. WAP spatial density.}
	\label{fig_ns_3_discovery}
\end{figure}

Fig \ref{fig_ns_3_discovery} shows the discovery performance of the discovery schemes. The y-axis values indicate the number
of discoveries per scan operation with observed RSSI greater than -75 dBm, for varying WAP spatial density. As expected from
the observation in the previous section, all the 
schemes show identical discovery performance. Hence, using the power consumption observations, 
it can be concluded that the WLAN-aware scheme
is the most energy efficient, as compared
to the other discovery schemes. Moreover, if the GPS coordinates of the UE are available beforehand, 
then the WLAN-aware-GPS results in the least number of
scanned channels for a mobile user, thereby, leading to small power consumption overhead.

\subsection{Power Consumption Modeling}
In this section, we provide a power consumption model for the WLAN-aware discovery scheme.
We assume that WAPs, operating in frequency $f_i \in F$, are distributed as Poisson Point process (PPP)
with spatial density given by $\rho_{f_i}$,
i.e., number of WAPs per m$^2$, and is independent of the distribution of WAPs operating on frequency $f_j \forall j \neq i$.
As such, the collective set of WAPs operating in any channel is also distributed as Poisson Point
with spatial density given by,
\begin{equation*}
	\rho = \sum_{i} \rho_{f_i}.
\end{equation*}
The consideration of WAP distribution as a PPP is supported by the measurement study conducted in \cite{Qiu}.

The average power consumption for a discovery scheme is provided in \cite{Wcnc}. As, power consumption values of \textit{Active Listen}
and \textit{Active Receive} state in many WLAN interfaces do not differ significantly, the expression of average power consumption
can be reduced to,
\begin{equation} \label{power}
	E[P_{scan}] = \cfrac {a \cdot E[N_{ch}]} {T_{s}} + b.
\end{equation}
Here, $N_{ch}$ corresponds to the number of channels scanned per scan operation and $T_{s}$ is the time interval to scan the selected channels. The value of $a$ depends
on the time interval to scan one channel and the power consumption in $Active Receive$ state, whereas, the value of $b$ corresponds to the 
power consumption in $Sleep$ state of the WLAN interface. The values of $a$ and $b$ can be determined
by observing the power consumption profile of scan operation for varying scan interval and number of channels to be scanned in each operation.
Hence, estimating the average power consumption of a discovery scheme corresponds to determining the expected value of $N_{ch}$.

%
%
%
%
%
%

\subsubsection*{Proposition 1}
Let the user be traveling with a speed of $v$ m/s. Let $T_s$ denote the scan interval. The expected number of channels scanned per scan
operation for the WLAN-aware scheme is given by,
\begin{equation} \label{nch}
	E[N_{wlan}] = \Big(N_1 + N_2 \cdot \cfrac{vT_{s}}{D_{1}} + N_3 \cdot \cfrac{vT_{s}}{D_{2}}\Big), 
\end{equation}

where,
\begin{equation} \label{n1}
	N_1 = \sum_{f_i \in F} (1 - e^{-\rho_{f_i} \pi D_{h}^2}),
\end{equation}
\begin{equation} \label{n2}
	N_2 = \sum_{f_i \in F} (1 - e^{-\rho_{f_i} \pi (D_{r}^2 - D_{h}^2)}) \cdot e^{-\rho_{f_i} \pi D_{h}^2},
\end{equation}
\begin{equation} \label{n3}
	N_3 = \sum_{f_i \in F} \Big(\int\limits_{\Delta=0}^{\infty}f_e(\Delta) (1 - e^{-\rho_{f_i} \pi((D_{r} + \Delta)^2 - D_{r}^2)})\Big) \cdot e^{-\rho_{f_i} \pi D_{r}^2}.
\end{equation}

Here, $f_e(\Delta)$ is the probability density function of the UE's estimated position error. The value of $D_{h}$ 
corresponds to the expected distance of a WAP from the UE, when, RSSI from the given WAP as observed by the UE is $S_{high}$.
The value can be determined using a known propagation loss model.

\subsubsection*{Proof}
Let $\Delta$ denote the estimated position error of the UE.
\emph{Initial Channel Selection} selects the operating channels of the WAPs that lie at distances less than $D_{r} + \Delta$,
from the current estimated position of the UE.
The UE scans the selected channels with different periodicity based on the previously observed RSSI values.
The criteria for selecting the periodicity of scanning each channel can be summarized as,

\begin{enumerate}
	\item If channel $f \in InitSelect$ satisfies $S[f] \geq S_{high}$, then, scan $f$ every $T_s$ interval,
	\item Else if channel $f \in InitSelect$ satisfies $S[f] < S_{high}$, then, scan $f$ after every $D_{1}/v$ interval,
	\item Else if channel $f \in InitSelect$ satisfies $S[f] = NaN$, then, scan $f$ after every $D_{2}/v$ interval.
\end{enumerate}
Using the above criteria, the expected number of channels scanned per scan operation is determined as,
\begin{equation} \label{nch}
	E[N_{ch}] = N_1 + N_2 \cdot \cfrac{vT_{s}}{D_{1}} + N_3 \cdot \cfrac{vT_{s}}{D_{2}},
\end{equation}
where, $N_1$, $N_2$ and $N_3$ represent the expected number of channels which satisfy Condition 1, 2 and 3, respectively.
Now, for PPP with WAP spatial density  given by $\rho$, the probability that at least one WAP is present in an area $A$ is given as,
\begin{equation} \label{general} 
	P[A] = 1 - \exp (-\rho A).
\end{equation}
Let, $d$ be the distance between the UE and a given WAP, and $S_{wap}$ be the RSSI observed by the UE from the given WAP. Since, the observed RSSI from a WAP decreases with increasing distance between
the UE and the WAP, the expression $S_{wap} \geq S_{high}$ can be approximated as $d \leq D_{h}$. The value of $D_{h}$ can be determined using a known
propagation loss model.
Therefore, using Equation \ref{general}, the value of $N_1$ is determined as,
\begin{equation} \label{n1}
	N_1 = \sum_{f_i \in F} (1 - e^{-\rho_{f_i} \pi D_{h}^2}).
\end{equation}

Using similar notion, the criteria $S_{wap} < S_{high}$ and the condition that a WAP is not detected by the UE, can be approximated
as $d \in (D_{h}, D_{r})$ and
$d \in (D_{r}, D_{r} + \Delta)$, respectively.
Again, using PPP, the probability that at least one WAP is deployed in an area $A_1$ such that no WAP is present
in area $A_2$, with $A_2 \subseteq A_1$ and the WAP spatial density $\rho$, is given by,
\begin{equation*}
	P = (1 - e^{-\rho (A_1 - A_2)}) \cdot e^{-\rho A_2}.
\end{equation*}
Using above, we can compute the values of $N_2$ and $N_3$ as,
\begin{equation} \label{n2}
	N_2 = \sum_{f_i \in F} (1 - e^{-\rho_{f_i} \pi (D_{r}^2 - D_{h}^2)}) \cdot e^{-\rho_{f_i} \pi D_{h}^2},
\end{equation}
\begin{equation} \label{n3}
	N_3 = E[\sum_{f_i \in F} (1 - e^{-\rho_{f_i} \pi((D_{r} + \Delta)^2 - D_{r}^2)}) \cdot e^{-\rho_{f_i} \pi D_{r}^2}].
\end{equation}

Combining Equations \ref{nch}, \ref{n1}, \ref{n2} and \ref{n3}, we can derive the expression for the expected number of channels scanned
for WLAN-aware scheme.

\subsubsection*{Proposition 2}
For the WLAN-aware discovery scheme, the probability density function of the estimated position error of the UE is given by,
\small
\begin{equation*}
	f_{e}(r) = \begin{cases} 
		2 \rho \pi r \cdot e^{-\rho \pi r^2} &  r\! \leq \! D_r, \\
		\delta(r - D_r - kvT_s) \cdot e^{-\rho \pi D_r^2} \big(\sum\limits_{i=k}^{\infty}\cfrac {P_{o}(i)}{i} \big) & r\!>\! D_r,k\!>\!0, 
	\end{cases}
\end{equation*}
\normalsize
where, $\delta(r)$ represents Dirac delta function and 
\begin{equation*}
	P_o(i) = e^{-2 D_r v \rho i T_s} (1 - e^{-2D_r v \rho T_s}).
\end{equation*}

\subsubsection*{Proof}
The position error of the UE is determined by the state of the UE in the context of WAP discoverability, i.e., $InRange$ and $OutRange$.
The UE is said to be in $InRange$ state, if there exists a WAP within $D_r$ distance from the UE. Thus, in $InRange$ state, the UE is able 
to receive signal from at least one WAP. The UE is said to be in $OutRange$ state, if no WAP exists within $D_r$ distance from
the UE and hence, the UE is not able to detect any WAP signal.
When the UE is in $InRange$ state, the estimated position error of the UE is given by the distance between the 
UE and the nearest WAP. Assuming time ergodicity and PPP, the probability density function of the distance 
between the UE and the nearest WAP can be derived as,
\begin{equation*}
	f(r) = 2 \rho \pi r \cdot e^{-\rho \pi r^2}.
\end{equation*}
Since, for $InRange$ state of the UE, the distance between the UE and the nearest WAP is less than $D_r$, the probability distribution
function for the estimated position error of the UE is given by,
\begin{equation}\label{lowErr}
	f_e(r) = 2 \rho \pi r \cdot e^{-\rho \pi r^2} \quad \text{if } r \leq D_r.
\end{equation}
When the UE enters $OutRange$ state, the initial estimated position error of the UE is $D_r$.
For the time UE stays in $OutRange$ state, the position error of the UE increments by $vT_{s}$ after each $T_{s}$ time interval.
Hence, if the UE stays in $OutRange$ state for $n$ consecutive scan intervals, the estimated position error of the UE is $D_r + kvT_s$ 
for $1/n \; \forall \; k \leq n$ fraction of time. 
Moreover, it has been shown in \cite{Kim} that the time of stay of UE in $OutRange$ state 
is exponentially distributed, i.e., $t_{o} \sim \exp(2D_rv\rho)$. Here, $v$ is the speed of the user
and $D_{r}$ is the transmission range of the WLAN. Hence, the probability of UE staying in $OutRange$ state for $n$ consecutive
scan intervals can be determined as,
\begin{equation}\label{timeOut}
	P_{o}(n) = e^{-2 D_r v \rho n T_s} (1 - e^{-2D_r v \rho T_s}).
\end{equation}
Since, the probability of UE being in $OutRange$ state is $e^{-\rho \pi D_r^2}$, the probability density function of the 
estimated position error of the UE is given by,
\small
\begin{equation}\label{highErr}
	f_{e}(r) =  \delta(r - D_r - kvT_s) \cdot e^{-\rho \pi D_r^2} \cdot \Big(\sum_{i=k}^{\infty}\cfrac {P_{o}(i)}{i} \Big) \quad \text{if }
	r >D_r, k> 0
\end{equation}
\normalsize
Using Equation \ref{lowErr}, \ref{timeOut} and \ref{highErr}, we can derive the expression for the probability density function 
of the estimated position error of the UE.

For the scenario of high WAP spatial density, $f_e(r)$ is expected to be high for small values of $r$, due to
more frequent WAP discoveries. On the other hand, for small values of WAP spatial density, the non-zero values of $f_e(r)$ are expected
to be evenly distributed over a wide range of values of $r$.

Using the proposition, it can be inferred that for low WAP spatial density scenario, the values of $N_1$ and $N_2$ are significantly small, but,
higher error in the position estimation of the UE leads to higher value of $N_3$. Hence, the discovery schemes, which use the
assistance of GPS module of the UE, are expected to have lower number of channel scans compared to WLAN-aware scheme.
As the WAP spatial density increases, higher accuracy in position estimation results in optimized selection of channels for $N_3$.
Consequently, the WLAN-aware scheme is expected to perform lower number of channel scans as compared to other discovery schemes.

Using similar methodology used for determining the expected number of channels scanned by the WLAN-aware scheme, we can determine the
expected number of channels scanned by the GPS-assisted scheme. The average number of channels scanned per scan operation for the GPS-assisted 
scheme is given by,
\begin{equation*}
       E[N_{gps}] = \sum_{f_i \in F} 1 - e^{(-\rho_{f_i} \cdot \pi D_{r}^2)}.
\end{equation*}
\begin{figure}[t]
	\centering
	\includegraphics[width=0.45\textwidth]{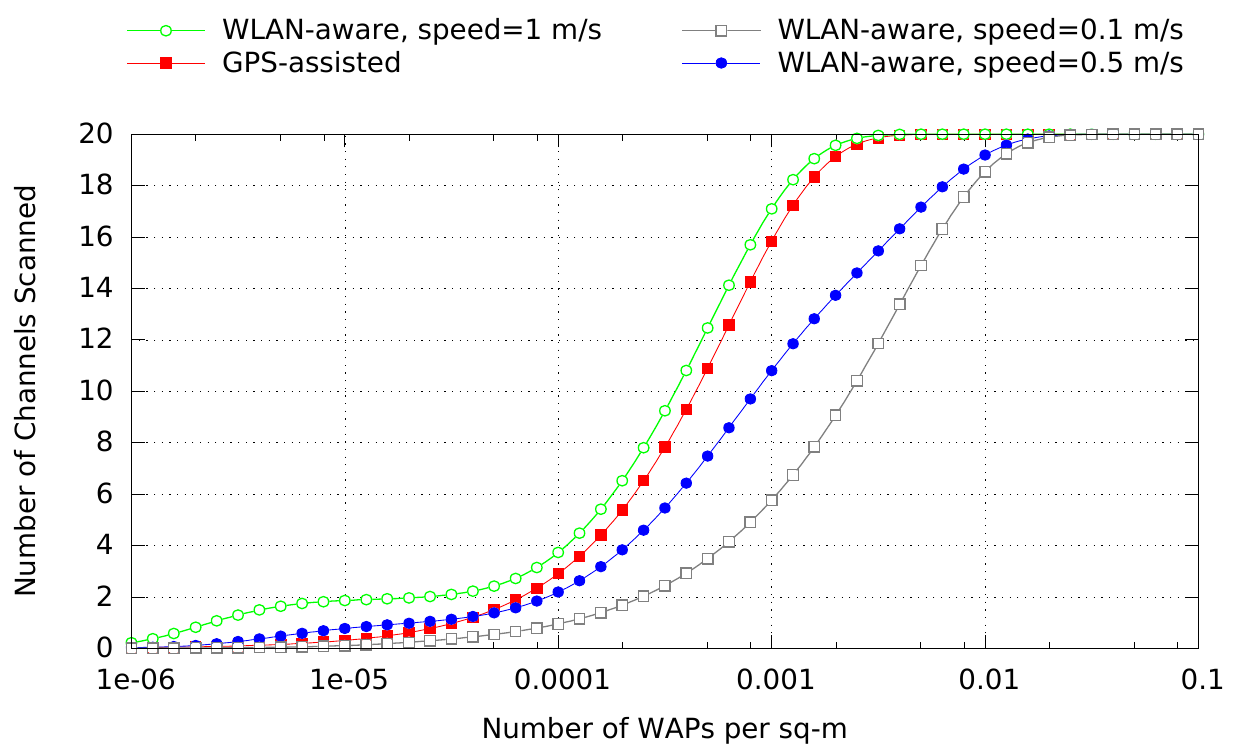}
	\caption{Number of channels scanned per scan operation as a function of WAP spatial density
	and user speed.}
	\label{fig_analysis_channels}
\end{figure}

Fig \ref{fig_analysis_channels} shows the trend for the number of channels scanned by the WLAN-aware scheme with varying WAP spatial density and
different user speeds. We have also plotted the number of channels scanned by the GPS-assisted scheme for performance comparison. 
It can be seen that the WLAN-aware scheme follows a similar trend as the results obtained from the simulation study.
The WLAN-aware scheme performs slightly higher number of channel scans as compared to the GPS-assisted
scheme, for low values of WAP spatial density.
As WAP spatial density increases, the WLAN-aware scheme shows similar 
performance to that of the GPS-assisted scheme, when user
speed is 1 m/s. Decreasing the user speed leads to significant performance improvement of the WLAN-aware scheme over the GPS-assisted scheme. 
This is because, the scan periodicity of channels corresponding to $N_2$ and $N_3$ is very low, for small values of user speed.
%
\section{Conclusions}
In this paper, we have proposed a WLAN-aware discovery scheme for heterogeneous 3GPP LTE networks. The WLAN-aware scheme differs
from the discovery
scheme implemented in most of the commercially available UEs, which periodically scans all the channels supported by the WLAN interface 
of the UE, by selecting the channels based on their
likelihood of providing offload opportunities. The scheme uses past scan results to evaluate the proximity
of WLANs deployed by the 3GPP network operator. We have implemented various discovery schemes along with the WLAN-aware scheme, on an Android phone, 
for performance evaluation. 
Further, we have also conducted simulation study in ns-3 and analytical modeling for comprehensive performance evaluation of the schemes.
Our results demonstrate that the WLAN-aware discovery scheme outperforms other solutions proposed in the literature, 
in terms of power consumption,
without compromising discovery performance. 
The field experimental and simulation results have been justified through the analytical model of power consumption also.
\section*{Acknowledgment}
This research work has been funded through a grant on LTE-WiFi Inter-working from Department of Electronics \& Information
Technology (DeitY), Government of India.
We thank Punit Rathod for his contributions towards the simulations presented in this paper.

\ifCLASSOPTIONcaptionsoff
  \newpage
\fi



\bibliographystyle{IEEEtran}
\bibliography{./bib1}
\end{document}